\documentclass{pasa}%

\def\kms{\,km\,s$^{-1}$}

\def\uJyperbeam{{\,$\upmu$Jy\,beam$^{-1}$}}
\def\pegase{{\sc p\'{e}gase}}
\def\mrmoose{{\sc MrMoose}}

\def\msun{\,M$_\odot$}
\def\mstel{\,$M_*$}
\def\fdg{.\!\!^\circ}
\usepackage{graphicx}
\def\lpco{\,{L$^\prime_{\rm CO}$}}
\def\lpcounit{\,K\,km\,s$^{-1}$\,pc$^{2}$}
\def\icounit{\,Jy\,km\,s$^{-1}$}
\def\mum{$\,\upmu$m}
\def\ghz{\,GHz}
\def\mhz{\,MHz}
\def\ks{$K_{\rm s}$}
\def\oddball{GLEAM\,J0917$-$0012}

\title[\oddball]{The Nature and Likely Redshift of GLEAM\,J0917$-$0012}
\author[Drouart et al.]{Guillaume Drouart$^1$\thanks{contact: guillaume.drouart@curtin.edu.au}, 
Nick Seymour$^1$,
Jess W. Broderick$^1$,
Jos\'e Afonso$^2$,
Rajan Chhetri$^{1,3}$,
Carlos De Breuck$^4$,
Bjorn Emonts$^5$,
Tim J. Galvin$^1$,
Matthew D. Lehnert$^6$,
John Morgan$^1$,
Daniel Stern$^7$,
Jo\"el Vernet$^4$, and
Nigel Wright$^1$
\affil{$^1$International Centre for Radio Astronomy Research, Curtin University, 1 Turner Avenue, Bentley, WA 6102, Australia}
\affil{$^2$Instituto de Astrof\'is\'ica e Ci\^encias do Espaço, Faculdade de Ci\^encias, Universidade de Lisboa, OAL, Tapada da Ajuda, PT1349-018 Lisboa, Portugal}
\affil{$^3$CSIRO Astronomy and Space Science, PO Box 1130, Bentley, WA 6102, Australia}
\affil{$^4$European Southern Observatory, Karl Schwarzschild Straße 2, 85748 Garching bei M\"unchen, Germany}
\affil{$^5$National Radio Astronomy Observatory, 520 Edgemont Road, Charlottesville, VA 22903, USA}%
\affil{$^6$Universit\'e Lyon 1, ENS de Lyon, CNRS UMR5574, Centre de Recherche Astrophysique de Lyon, 69230 Saint-Genis-Laval, France}%
\affil{$^7$Jet Propulsion Laboratory, California Institute of Technology, 4800 Oak Grove Drive, Pasadena, CA 91109, USA}%
}% 

\jid{PASA}
\doi{10.1017/pas.\the\year.xxx}
\jyear{\the\year}

\usepackage{aas_macros}
\usepackage{hyperref} 
\usepackage{upgreek}
\hypersetup{colorlinks,citecolor=blue,linkcolor=blue,urlcolor=blue}
\usepackage{mwe,tikz}\usepackage[percent]{overpic}
\usepackage{gensymb}
\usepackage[flushleft]{threeparttable}
\usepackage[T1]{fontenc}
\usepackage[final]{changes}
\makeatletter
\@namedef{Changes@AuthorColor}{red}
\colorlet{Changes@Color}{red}
\makeatother

\begin{document}

\begin{frontmatter}
\maketitle

\begin{abstract}
We previously reported a putative detection of a radio galaxy at $z=10.15$, selected from the GaLactic and Extragalactic All-sky Murchison Widefield Array (GLEAM) survey. The redshift of this source, GLEAM\,J0917$-$0012, was  based upon three weakly detected molecular emission lines observed with the Atacama Large Millimetre Array (ALMA). In order to confirm this result, we conducted deep spectroscopic follow-up observations with ALMA and the Karl Jansky Very Large Array (VLA). The ALMA observations targeted the same CO lines 
\deleted{we had} previously reported in Band 3 (84-115\,GHz) and the VLA targeted the CO(4-3) and [CI(1-0)] lines for an independent confirmation in $Q$-band (41 and 44\,GHz). Neither observation \added{detected any}\deleted{found} emission lines, \added{removing}\deleted{taking away} support for our original interpretation. \added{Adding publicly available optical data from the Hyper Suprime-Cam survey, {\it WISE} and {\it Herschel Space Observatory} in the infrared, as well as $<$10\,GHz polarisation and 162\mhz\ inter-planetary scintillation observations, we model the physical and observational characteristics of \oddball\ as a function of redshift. Comparing these predictions and observational relations to the data\deleted{and information}, we \added{are able to} constrain \deleted{both} its nature and distance.} \deleted{Using all available multi-wavelength data, including further data from the literature,} We argue that if GLEAM\,J0917$-$0012 is at \replaced{$z<3$}{$z<7$} then it has an extremely unusual nature, and that the more likely solution is that the source lies above $z=7$.
\end{abstract}

\end{frontmatter}

\section{INTRODUCTION }
\label{sec:intro}

The discovery of a legion of supermassive black holes (SMBHs; $\sim$ $10^8\,$\msun) at high redshift \citep[$z>6$;][]{banados_800-million-solar-mass_2018} has intensified the discussion on how such objects can form \added{and grow} so quickly \citep[e.g.][]{volonteri_rapid_2005}. \added{Samples of a}ctive galactic nuclei (AGN) \deleted{samples} are selected \added{via indicators} across the electromagnetic spectrum in X-rays, optical, near-infrared (near-IR), mid-IR or radio, all of which originate from one of several \replaced{different physical}{of the} manifestations \deleted{of the AGN} as described by the \added{AGN} unification scheme \citep[e.g.][]{antonucci_unified_1993}. While relatively large and deep X-ray surveys \replaced{will soon be within}{are just in} reach with the extended ROentgen Survey with an Imaging Telescope Array \citep[eROSITA;][]{cappelluti_erosita_2011},  recent years have seen \replaced{increasing numbers of}{more and more} AGN detected at $z>5$ \citep{banados_pan-starrs1_2016} due to the proliferation of deep, large optical and near-IR surveys \citep[e.g. Panoramic Survey Telescope and Rapid Response System; Pan-STARRS and VISTA Kilo-Degree Infrared Galaxy Survey, VIKING;][]{arnaboldi_eso_2007,chambers_pan-starrs1_2016}. The mid-IR selection \replaced{has provided}{showed} some successful selection criteria \citep[e.g.][]{stern_mid-infrared_2005}, but is currently stalled \replaced{with}{given that} no large \added{mid-}IR surveys \deleted{are} foreseen in the near future. As for radio-selected samples, \replaced{these}{they} have been lagging due to the lack of deep and wide low-frequency surveys which allow\deleted{s} one to efficiently isolate high-redshift candidates among \added{the} millions of \added{radio} sources already \replaced{catalogued}{present in large radio surveys} \citep[e.g.][]{de_breuck_sample_2000}. With the release of the TIFR Giant Metrewave Radio Telescope (GMRT) Sky Survey \citep[TGSS;][]{intema_gmrt_2017}, the GaLactic and Extragalactic All-sky Murchison Widefield Array \citep[MWA; ][]{tingay_murchison_2013} survey \replaced{\mbox{\citep[GLEAM;][]{wayth_gleam:_2015}}}{{\mbox{\citep[GLEAM;][]{hurley-walker_galactic_2017}}}}, the LOFAR Two-metre Sky Survey \citep[LoTSS;][]{shimwell_lofar_2017,shimwell_lofar_2019}, and more recently the Rapid Australian Square Kilometre Array Pathfinder (ASKAP) Continuum Survey \citep[RACS;][]{mcconnell_rapid_2020}, we finally have the opportunity to continue the search for \replaced{radio-selected}{radio-loud} AGN at high redshift.

% current record holders - why radio? 
The most distant optically-selected AGN currently known is at $z=7.64$ \citep{wang_luminous_2021}, while the most distant radio-loud object is at $z=6.82$ \citep{banados_discovery_2021}\footnote{\added{Although this radio source is relatively weak in the radio and only just qualifies as radio-loud (it is $1-2$ orders of magnitude less bright at 1.4\,GHz than the most distant radio-selected AGN).}}. Both optically\added{-selected} and radio-selected samples are complementary and necessary to capture a complete picture of \added{early} \deleted{the} AGN evolution within the Epoch of Reionisation (EoR). While optical selection tends to miss dust-obscured objects (due to obscuration by the host galaxy and/or the orientation of the torus with respect to the observer), radio selection alleviates this bias, at the cost of only capturing the radio\replaced{-luminous}{-loud} population.

% radio galaxies in particular
\added{
Powerful high-redshift radio galaxies \citep[HzRGs; $L_{3\,\rm GHz}>10^{26}\,$WHz$^{-1}$;][]{miley_distant_2008} have a long history of being AGN/galaxy distance `record holders' \cite[e.g.][]{van_breugel_radio_1999}. These advances were a result of a number of wide-area radio surveys available in the 1980s to 1990s and the advent of the ultra-steep spectrum (USS) technique \citep[e.g.][using the radio spectral index for selection, $\alpha\le -1.3$; $S_\nu\propto\nu^\alpha$]{de_breuck_sample_2000}. Radio selection of AGN typically includes faintness in the $K$-band as powerful radio galaxies follow the $K-z$ relation \citep[e.g.][]{rocca-volmerange_radio_2004}. As they are highly obscured AGN, the rest-frame optical emission from these sources is dominated by a massive stellar population \citep[$\sim 10^{11}\,M_\odot$;][]{seymour_massive_2007}  leading to a correlation of $K$-band emission with redshift.

The recent discovery of the USS-selected source TGSS J1530$+$1049 at $z=5.72$ \citep[$\alpha=-1.4$; ][]{saxena_discovery_2018} broke the 20-year old radio-powerul AGN distance record from \cite{van_breugel_radio_1999}. We have developed a new selection criterion based upon the low-frequency ($70-230\,$MHz) curvature of a radio source's spectral energy distribution (SED) in GLEAM \citep[][hereafter D20]{drouart_gleaming_2020}. This curvature is due to the expected small size of radio-loud AGN at $z>5$ \citep[e.g.][]{saxena_modelling_2017} and the presence of a low-frequency turn-over due to synchrotron self-absorption and/or free--free absorption processes.  This method is agnostic to the nature of the high-frequency SED, and hence finds sources which do not have ultra-steep spectra. However, it does require the same faintness in $K-$band as the USS selection (in this case \ks$>21.2$ in VIKING) to further refine the identification of high-redshift candidates. Using this selection method, D20 discovered GLEAM\,J0856+0224 at $z=5.55$, which has a spectral index of $\alpha\sim -0.78$ across $0.1-1\,$GHz and would have been missed by the USS selection technique.

Very recently, a handful of optically selected quasi-stellar objects (QSOs) known at $z>6$ have been detected in radio surveys: at $z=6.44$ \citep{ighina_radio_2021} and $z=6.82$ \citep{banados_discovery_2021}. However, these radio-loud QSOs are more than an order of magnitude less luminous than HzRGs in the radio (not meeting the 3-GHz rest-frame luminosity threshold) and are potentially beamed as both show evidence of radio variability.}

\deleted{Powerful high-redshift radio galaxies \mbox{\citep[HzRGs;][]{miley_distant_2008}} have a long history of being distance `record holders' as a result of the extensive number of radio surveys made available in the 1990s and the ultra-steep spectrum (USS) selection technique \mbox{\citep[e.g.][]{de_breuck_sample_2000}}. With the recent discovery of the USS source TGSS J1530$+$1049 at $z=5.72$ \mbox{\citep[][]{saxena_discovery_2018}}, as well as our discovery of GLEAM\,J0856+0224 at $z=5.55$ using a new HzRG selection technique \mbox{\citep[][D20 hereafter]{drouart_gleaming_2020}}, the 20-year old distance record of $z=5.19$ for HzRGs \mbox{\citep[TN J0924$-$2201;][]{van_breugel_radio_1999}} was broken, paving the way for a new generation of samples at even higher redshift. Due to their preferential orientation in the plane of the sky \mbox{\citep{drouart_jet_2012}}, HzRGs present a prime opportunity to simultaneously study the SMBH, host galaxy, and local environment, as well their co-evolution \mbox{ \citep[e.g.][....]{seymour_massive_2007,nesvadba_black_2011,drouart_rapidly_2014,wylezalek_galaxy_2014,emonts_co1-0_2014,drouart_disentangling_2016,falkendal_massive_2019}}.}

% the case of 0917
\deleted{In D20, we described a pilot study on a new technique to select HzRGs at $z>5$.} \added{In D20, we also} presented a tentative $z=10.15$ redshift for GLEAM~J091734-001243 (hereafter \oddball), based on the presence of three low signal-to-noise, putative carbon monoxide (CO) lines in an Atacama Large sub-Millimetre Array (ALMA) $84-115\,$GHz spectrum. The CO lines were extracted at the position of the host galaxy detected in a deep \ks-band image from the HAWKI instrument on the Very Large Telescope (VLT). In this paper, we describe supplementary ALMA and Karl G. Jansky Very Large Array (VLA) data that do not confirm this extreme redshift. However, we present arguments on the high-redshift nature of \oddball\, using a compilation of multi-wavelength data. The paper is organised as follows. In Section~\ref{sec:data}, we present our new ALMA and VLA follow-up observations, as well as additional data \added{at radio, IR and optical wavelengths} from the literature. Section~\ref{sec:results} shows the constraints on the \deleted{spectral energy distribution} SED of  \oddball\, in the optical/near-IR and radio/far-IR. In Section~\ref{sec:z_0917}, we attempt to constrain \added{the nature of \oddball\ with respect to redshift, jointly using model predictions \replaced{,}{and} empirical relations and observational limits.} \deleted{a possible redshift for \oddball\, using these data.} We then discuss the likely \added{range of} redshifts of \oddball\, in Section~\ref{sec:disc} \added{when combining all the information derived in the previous sections}, before \replaced{encapsulating}{reporting} our conclusions in Section~\ref{sec:conc}. Throughout this paper, we assume a flat $\Lambda$CDM cosmology with H$_0$=67.7 km\,s$^{-1}$\,Mpc$^{-1}$ and $\Omega_M$=0.308 \citep{planck_collaboration_planck_2016}.

% NS: just brain-storming the motivation OK here:\\
% - why high-z RL are important\\
% - work on high-z AGN to date and different techniques\\
% - explain aim of this paper: D20 found a z=10 candidate, we wanted to confirm with VLA and ALMA. \\
% - mention selection technique and reference GLEAM and VIKING \\
% - did not confirm original lines so what is this source?\\
% - if not at $z>x$ then very unusual, need to discuss how we selected \oddball and what we found. It is quite extreme in $K-z$ and $S_{\rm 1.4\ghz}/S_K$\\

% \begin{itemize}
%     \item can pump shamelessly on the proposal scientific rationale. (NS: but shorten, see my suggested structure above)
%     \item need to mention the Hashimoto source, REBELS, and some other ALMA high-z detection. 
% \end{itemize}

\section{DATA}
\label{sec:data}

Following our previous observing campaign presented in D20, we further observed \oddball\, in order to confirm or refute our original tentative redshift determination: $z=10.15$. The aim of our VLA observation in $Q$-band ($36-46\,$GHz) was to independently confirm the redshift by the detection of the CO(4-3) and \added{carbon} [CI(1-0)] lines, whereas the deeper ALMA observations aimed to  confirm the original CO(9-8), CO(10-9) and CO(11-10) line detections. We also obtained further data from the literature in order to compile a well-sampled broadband SED, thereby allowing us to estimate source properties.

\subsection{\added{Host identification}}
\begin{figure*}
    \centering
    \includegraphics[width=0.95\textwidth, trim= 0 0 0 0,clip]{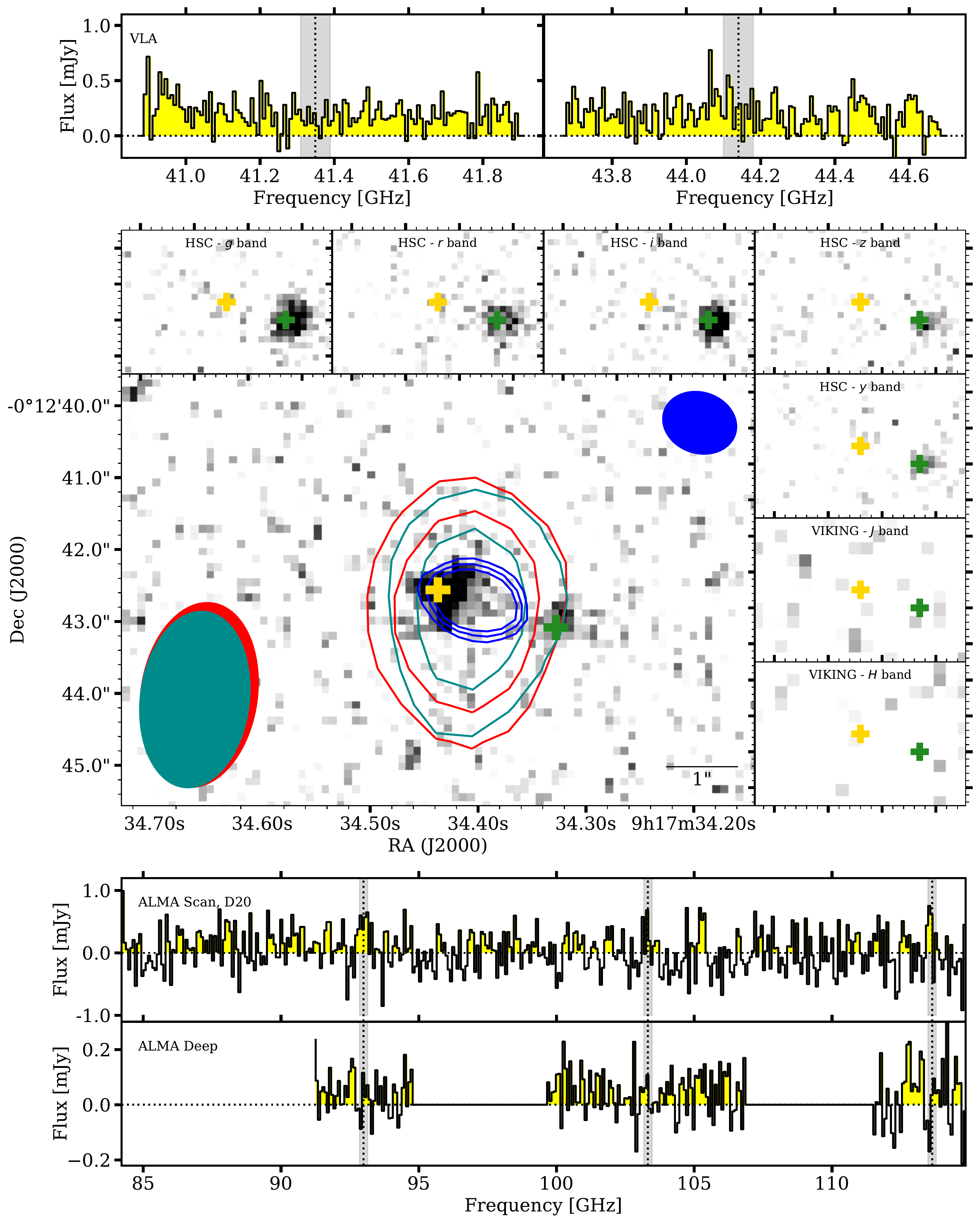}
    \caption{{\it Top:} VLA spectra extracted at the host galaxy position with dotted lines/grey regions indicating expected locations of the targeted CO(4-3) and [CI(1-0)] lines. {\it Middle:} $K_s-$band image \added{(centre)} and the available HSC and VIKING data in insets in greyscale, with the \replaced{deepest}{new} ALMA continuum image (blue contours) \added{at 3, 4 and 5$\sigma$} and the respective VLA continuum \added{41- and 44-GHz} images (red and dark-cyan contours) overlaid at 5 and 10$\sigma$. The beams are presented in the corners in their respective colours. The yellow cross indicates the coordinates of the host galaxy used to extract the presented spectra. \added{The green cross on the detected source south-west of the host represents the coordinates for the aperture photometry presented in Fig.~\ref{fig:SW-source}. Note that the cross are $\sim$0.3 arcsec wide, corresponding to the absolute positional accuracy of our data (see Section~\ref{sec:data}).} {\it Bottom}: \added{ALMA spectra from D20 (top) and the new, deeper follow-up spectrum (bottom), extracted at the host galaxy position (yellow cross). Note the change in flux scale ($\times4$) and the dotted line/grey regions indicating expected locations of the targeted CO(9-8), CO(10-9) and CO(11-10) lines}.\deleted{ In the top and bottom panels, the dashed lines and grey shaded areas represent the expected position for the CO lines for a source at $z=10.15$.}}
    \label{fig:allradio}
\end{figure*}

\added{
The absolute astrometric uncertainties of the HAWKI, ALMA and VLA data are small (0.2-0.3\,arcsec, see the respective following subsections); hence we have reliably identified the host galaxy responsible for the radio emission as the \ks-band source (yellow cross in Fig.~\ref{fig:allradio}) within the 100-GHz beam, but 0.5\,arcsec east of the ALMA central coordinates. Some offset between the ALMA and the \ks-band coordinates is not unexpected due to the nature of the emission; the former from the synchrotron emission from the jets and the later from the stellar emission from the host galaxy.

The second source visible $\sim$1\,arcsec south-west of the ALMA 100-GHz continuum (green cross in Fig.~\ref{fig:allradio}) is believed to be unrelated to the host galaxy. We present its SED in Section~\ref{sec:optical_SED}. We also extracted the ALMA and VLA spectra at these coordinates (following the same method presented in the next subsections) and identified nothing (lines or continuum) to report.}

\subsection{VLA data}

Our DDT programme was observed by the VLA in D-configuration on 2020 January 6 and 7 (ID: 19B-337). The  time on-source was 5\,h (total of 10\,h including overheads) in $Q$-band. We set up the 8-bit correlator in order to (i) simultaneously cover the redshifted CO(4-3) and [CI(1-0)] lines (observed-frame frequencies 41.33\ghz\ and 44.12\ghz, respectively, assuming $z=10.15$), (ii) place each of these lines at the centre of a 128-MHz subband to optimise the signal-to-noise ratio, and (iii) perform a 20-MHz offset between the two observations to fill the `gaps' at the edges of the 128-MHz subbands. The final data therefore had continuous frequency coverage of the two lines, with a relatively homogeneous noise level in the 1-GHz bandwidth centred on each line. We performed calibration from the raw data using {\sc casa} v5.6.2-2 \citep{mcmullin_casa_2007} in the pipeline mode with default values. We visually checked the visibilities and performed extra flagging as required (i.e. one antenna was flagged due to noisy amplitudes). 

% \begin{figure*}[t]
%     \centering
%     \includegraphics[width=1.0\textwidth]{VLA_spectra.pdf}
%     \caption{{\it top:} VLA spectra extracted on the host coordinates from the $K_s-$band image (see D20), symbolised by an open circle. {\it bottom:} The greyscale images represent the continuum emission at 41.33\,GHz {\it (left)} and 44.12\,GHz {\it (right)}. The contours (TBD) represent the continuum-subtracted flux integrated over the grey regions of the spectrum (centred on the frequency for the expected redshift of CO(4-3) and[CI(1-0)] respectively). \warning{Need to add the contours corresponding to the grey shaded area+cosmetics}. Note: the uniform-weighted images show RFI, looking into this at the moment. No emission lines are observed in these data.}
%     \label{fig:spectra_VLA}
% \end{figure*}

\subsubsection{Continuum image}

Creating images with natural weighting, the synthesised beams are 2.6\,arcsec$\times 1.7$\,arcsec (FWHM with beam position angle PA=$-9\degree$ measured north through east) and 2.5\,arcsec $\times 1.6$ arcsec (PA$=-8\degree$) at 41.4 and 44.2\ghz, respectively (see Fig.~\ref{fig:allradio} lower panel). The one sigma sensitivity levels are 13 and 15\,\uJyperbeam\, at 41.4 and 44.2\ghz, respectively\added{, with an absolute positional accuracy}\footnote{\url{https://science.nrao.edu/facilities/vla/docs/manuals/oss/performance/positional-accuracy}}\added{ of 0.2\,arcsec}. The source is well detected at the location of the host galaxy in continuum in the two 1-GHz frequency bands and appears unresolved \added{(see Fig.~\ref{fig:allradio})}. We extracted the flux density \added{in the continuum images} at each frequency using {\sc aegean}\added{, a source finding and photometry code performing 2D-Gaussian fits on detected sources in an image} \citep{hancock_source_2018}. The \replaced{integrated flux densities}{results} are reported in Table~\ref{tab:radio_flux}.

\subsubsection{Spectral data cube}

Using natural weighting and applying an 8-MHz channel width, the data cubes reach sensitivities of 120 and 150\uJyperbeam at 41.4 and 44.2\ghz, respectively. We present the resulting spectra in Fig.~\ref{fig:allradio}, extracting at the host galaxy coordinates \added{(the yellow cross), following the same procedure described in D20, i.e. an average spectrum assuming a 0.8-arcsec aperture}. While the continuum is clearly detected, the expected lines are not detected. We further explore the implications of the non-detections in Section~\ref{sec:z_0917}. 
 
\subsection{Deeper ALMA data}

Our DDT programme was observed in C43-4 configuration on 2020 March 14 and 15 (ID:2019.A.00023.S). The observations consist of two tunings in Band 3, centred at 99.0 and 107.4\ghz, in order to cover the three CO transitions with enough channels surrounding each line for a reliable continuum fit. All data analysis was performed with {\sc casa} v5.6.2. \added{In particular, one should note the difference between the original ALMA dataset covering a 30-GHz bandwidth with five tunings of $\sim$10\,min, and the new, deeper follow-up consisting of two 7.5-GHz-bandwidth tunings, each with an on-source integration time of $\sim$40\,min. This results in the new continuum images reaching a similar continuum depth if taken individually and compared to the previous ALMA data from D20. However, it is important to note that the line sensitivity is increased by a factor of $\geq 2$.}

\subsubsection{Continuum image}

Given the $\sim$40-min on-source time for each tuning, we imaged each of the two tunings separately to obtain continuum detections at both 99.0\,GHz and 107.4\,GHz. We use natural weighting in order to optimise the sensitivity. The resulting flux densities, measured \added{in the continuum images} with {\sc aegean} and reported in Table~\ref{tab:radio_flux}, are consistent with the collapsed 30-GHz-bandwidth flux density presented in D20. The final sensitivities are 11 and 14\uJyperbeam~at 99.0 and 107.4\ghz, respectively; the synthesised beams are 1.41\,arcsec $\times 1.24$\,arcsec (PA$=-69\degree$) at 99.0\ghz, and 1.46\,arcsec $\times 1.17$\,arcsec (PA$=76\degree$) at 107.4\ghz~\added{(similar to the resolution obtained in D20) and the positional accuracy}\footnote{See section 10.5.2 of the ALMA technical handbook, \url{https://almascience.eso.org/documents-and-tools/}.}\added{ reaches $\sim$0.3\,arcsec}. As such, we now have three independent data points with which to investigate the 100-GHz part of the SED (discussed further in Section~\ref{sec:radio_SED}). \added{We also create a continuum image concatenating all visibilities (the 30\,GHz and the two 7.5\,GHz data cubes) to generate the best resolution map. The optimal sensitivity and resolution are 6.3\,$\upmu$Jy and 1.04\,arcsec $\times 0.86$\,arcsec (PA$=70\degree$) arcsec, using a Briggs parameter of 1, presented in Fig.~\ref{fig:allradio} as blue contours.}

\subsubsection{Spectral data cube}

We concatenated the visibilities from the two new datasets and imaged the data cube. First, we used natural weighting, and then a 3-arcsec tapered beam using 80-MHz-width channels to be consistent with our previous observations from D20. The final cubes reach average noise levels of 70 and 140\uJyperbeam~per 80\,MHz-width channel, respectively. Fig.~\ref{fig:allradio} presents the spectra extracted at the host galaxy coordinates from the \ks-band image, \added{following the same procedure described in D20 by averaging the spectra over an 0.8-arcsec aperture}, for the previous and new 3-arcsec tapered data. None of the previous lines are confirmed, and we therefore do not confirm our previous tentative $z=10.15$ solution. The implications of this are further discussed in Section~\ref{sec:z_0917}. An extraction at the peak of the radio continuum or at the position of the second source detected in the \ks-band image did not reveal any lines.

\begin{table}[t]
    \centering
    \caption{Continuum flux densities and their respective uncertainties for each image from both publicly available data and our new VLA and ALMA data. \added{Uncertainties include a 10\% calibration uncertainty added in quadrature for all radio fluxes, excepted for the RACS flux which follows the Equation~7 from} \citet{mcconnell_rapid_2020}). The reported upper limits are at the $3\sigma$ level. The GLEAM, TGSS and NVSS data are not included here; Table~4 in D20. References: [M13] \citet{mauch_325-mhz_2013}; [B95] \citet{becker_first_1995}; [M20] \citep{mcconnell_rapid_2020}; [G20] \citet{gordon_catalog_2020}; [TP] this paper; [D16] \citet{driver_galaxy_2016}.}
    \begin{tabular}{l c c c}
        \hline
        Facility/band & Freq. [GHz] & Flux [mJy] & Ref \\
        \hline \hline
        GMRT & 0.325 & 277$\pm$28 & [M13] \\
        RACS & 0.887 & 83.8$\pm$6.4 & [M20] \\
        FIRST & 1.4 & 47.3$\pm$4.7 & [B95]\\
        VLASS & 3.0 & 16.1$\pm$1.6 & [G20]\\
        VLA\_CO & 40.9 & 0.29 $\pm$ 0.044 & [TP] \\ 
        VLA\_CI & 43.7 & 0.24 $\pm$ 0.031 & [TP] \\ 
        ALMA\_B3 & 100 & 0.060 $\pm$ 0.013 & [TP] \\ 
        ALMA\_Deep1 & 99 & 0.067 $\pm$ 0.017 & [TP] \\ 
        ALMA\_Deep2 & 107 & 0.078 $\pm$ 0.020 & [TP] \\ 
        \hline
        SPIRE\,500\mum & 600 & <9.9 & [D16] \\ 
        SPIRE\,350\mum & 857 & <8.2 & [D16] \\ 
        SPIRE\,250\mum & 1,200 & <6.7 & [D16] \\ 
        PACS\,160\mum & 1,880 & <20 & [D16] \\ 
        PACS\,100\mum & 3,000 & <18 & [D16] \\ 
        {\it WISE} 22\mum & 13,600 & <0.48 & [D16] \\ 
        {\it WISE} 12\mum & 25,000 & <0.08 & [D16] \\ 
        {\it WISE} 4.5\mum  & 66,700 & <0.016 & [D16] \\ 
        {\it WISE} 3.6\mum  & 83,300 & <0.008 & [D16] \\ 
         \hline
    \end{tabular}
    \label{tab:radio_flux}
\end{table}

%hsc_flux_limit_from_fits=np.array([8.0e-8,1.2e-7,1.35e-7,2.32e-7,3.15e-7])
%3.069645542037043 0.12020684675709052
%[0.6186003  1.49437965 1.80934136 3.92587034] - wavelength=np.array([0.88,1.02,1.25,1.64,2.14])
% # G: mag AB 26.64; flux 8.00e-08 Jy
% # R: mag AB 26.20; flux 1.20e-07 Jy
% # I: mag AB 26.07; flux 1.35e-07 Jy
% # Z: mag AB 25.49; flux 2.32e-07 Jy
% # Y: mag AB 25.15; flux 3.15e-07 Jy
% hsc_wav=np.array([0.4754,0.6175,0.7711,0.8898,0.9762])

\subsection{Inter-planetary scintillation observations}
\label{sec:IPS}

The GAMA-09 field was covered by observations with the MWA, searching for sub-arcsecond-scale structures in sources using the phenomenon of interplanetary scintillation (IPS) at 162\mhz. IPS arises due to turbulence and structure in the solar wind, causing the radio emission from angular scales \replaced{at $\lesssim 0.3$\,arcsec (Fresnel size for IPS at 162\mhz)}{ of less than$\sim 0.5''$} to scintillate on timescales of $\sim$1 second. Using MWA wide-field images with a 0.5-s cadence over 10-min observations, the RMS flux density variation of each bright source is used to calculate its normalised `scintillation index'  \citep[NSI; ][]{morgan_interplanetary_2018,chhetri_interplanetary_2018}. The NSI can be used as an estimator of angular size, or of the flux arising from the compact component. \oddball\, was found to have a significant median NSI of 0.49$\pm$0.03 from 29 such observations. \added{This value implies one of the three following scenarios \citep[as discussed in Fig.~5 of ][]{morgan_interplanetary_2018} for source morphology.} 
\begin{itemize}
    \item \added{A slightly resolved Gaussian, here
    approximately twice the size of the Fresnel diameter, i.e. $\sim0.6$\,arcsec.}
    \item \added{A point source with half of the total flux embedded in an extended component (which could in theory be as large as the MWA synthesised beam at this frequency). }
    \item \added{Two compact components separated from each other by $>0.3$\,arcsec where one is partially resolved at $\sim0.3$\,arcsec.}
\end{itemize}
\added{Note that the second and third scenarios cannot provide us with an upper limit for the spatial scale of the extended component. With help of other radio data presented in this paper, }
\deleted{, suggesting around half its flux density at 162\mhz\, is on a scale $\leq 0.5$\,arcsec.} we further discuss the implication of this size \added{and morphology} in Section~\ref{sec:z_0917}. 

\subsection{Radio polarimetric properties at $<$\,10\,GHz}
\label{sec:polarisation}

In order to gather as much information as possible to discuss the nature of \oddball\ (Section~\ref{sec:disc}), we checked for radio polarisation at $\nu<$10\ghz, where our signal-to-noise is the most suitable to perform this analysis. In linear polarisation, \oddball\, is detected neither in the $169$--$231$ MHz MWA POlarised GLEAM Survey \citep[POGS; $7\sigma$ fractional polarisation upper limit $\approx 4.2$ per cent for \oddball;][]{riseley_polarised_2020} nor in the $1.4$-GHz NRAO VLA Sky Survey \citep[NVSS; $5\sigma$ fractional polarisation upper limit $\approx 4.7$ per cent for \oddball;][]{condon_nrao_1998}. We also investigated the radio polarimetric properties from our $5.5$- and $9$-GHz ATCA data (D20). We first conducted further processing of the data in Stokes $I$: we ran phase-only self-calibration on both the $5.5$- and $9$-GHz data sets. The flux densities at both frequencies remained consistent with the values previously reported in D20. We then imaged the combined $5.5$$+$$9$-GHz dataset so as to create a map at $7.25$ GHz. Using a robust weighting parameter of $0.5$, the angular resolution is 43.4\,arcsec $\times$ $21.8$\,arcsec (PA=$69\fdg8$). \oddball\, has peak and integrated $7.25$-GHz flux densities of $6.0 \pm 0.6$ mJy beam$^{-1}$ and $6.1 \pm 0.6$\,mJy, respectively.     

A $7.25$-GHz Stokes-$V$ map was then constructed using the same imaging settings as above. The RMS noise level in this map is $31$ $\upmu$Jy beam$^{-1}$. \oddball\, is not detected in Stokes $V$, with the $5\sigma$ upper limit for the fractional circular polarisation at $7.25$\,GHz being approximately $2.5$ per cent. 

To investigate the linear polarimetric properties at $7.25$ GHz, we used the {\sc rm-tools} software package \citep{purcell_rm-tools_2020} to conduct Faraday synthesis. Stokes $I$, $Q$ and $U$ images were made for all $3513 \times 1$-MHz channels from $4.577$--$9.923$ GHz that had not been flagged. Again, we used a robust weighting parameter of $0.5$; moreover, all images had the same pixel size ($5$ arcsec) and were restored with the same synthesised beam (taking the coarsest angular resolution from the lowest-frequency channel: 59.0\,arcsec $\times$ $38.6$\,arcsec and PA $= 71\fdg4$). We then generated Faraday depth spectra, corrected for the slope of the Stokes-$I$ in-band spectrum, for each pixel in a $3 \times 3$ grid centred on the pixel of peak intensity in Stokes $I$. No components above $5\sigma$ were found in any of the Faraday depth spectra. Similar to the result for the fractional circular polarisation, the $5\sigma$ upper limit for the fractional linear polarisation is approximately $2.6$ per cent (the average full-bandwidth noise level in Stokes $Q$ and $U$ is $32$ $\upmu$Jy beam$^{-1}$).   

In conclusion, \oddball\, has not been detected in polarisation yet at metre or centimetre wavelengths. \added{We discuss the implication in Section~\ref{sec:disc}.}
%Further polarimetric analysis, including using our ATCA $17$- and $19.4$-GHz data (D20), is beyond the scope of this paper.  

%Although \oddball is not associated with a known pulsar\footnote{There are no known pulsars with in 1\,degree in the ATNF Pulsar Catalogue: www.atnf.csiro.au/research/pulsar/psrcat/. Also it is not detected in MWA pulsars surveys to date (R. Bhat, private communication).}, compact radio sources with steep spectra and significant polarisation have been followed up and subsequently found to be Galactic pulsars in previous studies \citep[e.g.][]{navarro_very_1995}. However, \oddball was not detected in linear polarisation in the $1.4$-GHz NVSS survey \citep[$5\sigma$ fraction polarisation upper limit $\approx 4.7$ per cent;][]{condon_nrao_1998}, nor in the $169$--$231$ MHz MWA POlarised GLEAM Survey \citep[POGS;][]{riseley_polarised_2020}. Furthermore, there is no evidence in the radio spectrum of variability, which we might expect from a pulsar, particularly when the spectrum has been compiled using data from a variety of observations taken at different epochs. Additionally, the optical and near-IR (near-IR) properties do not particularly suggest a pulsar origin, although, for example, some pulsars have been detected at near-IR wavelengths before \citep[e.g.][]{mignani_near-infrared_2012}. Hence this combination of radio and near-IR properties does not provide compelling evidence at all that \oddball is a nearby pulsar.     

%\warning{(JB: we need to check if Ramesh searched the beamformed MWA data as well)}

\subsection{Literature data}

We also compiled publicly available data on \oddball, from optical to radio wavelengths. After visual confirmation of a non-detection in the images, we report additional upper limits from the {\it Herschel Space Observatory} \citep[\added{hereafter {\it Herschel}}][]{pilbratt_herschel_2010}, the {\it Widefield Infrared Survey Explorer} \citep[{\it WISE};][]{wright_wide-field_2010} and the Hyper Suprime-Cam Subaru Strategic Program \citep[HSC;][]{aihara_first_2018}. The HSC survey covers our source with the {$grizy$} bands, reaching significantly deeper than the VIKING survey ($zYJHK$) in the overlapping bands, thereby allowing for stronger constraints to be placed on the optical SED. We downloaded the HSC images and performed aperture-matched photometry at the \replaced{source coordinates}{host}  based on the detection and the resolution in \ks-band ($\sim$0.7\,arcsec; see D20). The flux densities and associated uncertainties are given in Table~\ref{tab:optical_flux} along with all the limits from the VIKING survey and our previously reported $K_s-$band detection (D20). 

In the case of the mid- and far-IR flux densities, we assumed that the source is unresolved and report the {\it Herschel} and {\it WISE} survey limits provided in \citet{driver_galaxy_2016} in Table~\ref{tab:radio_flux}, along with new radio data obtained from the literature. 

\begin{table}[]
    \centering
    \caption{Optical and near-IR limits based on the $K_s-$band detection \added{ for \oddball\ and the south-western (SW) source seen in \ks-band.} We present the SED filter, central wavelength ($\lambda_0$) and flux densities ($F_{\nu}$). The upper limits are at the 3$\sigma$ level from aperture-matched photometry using the \ks-band detection. These data are plotted in Fig.~\ref{fig:optical_SED}.}
    \begin{threeparttable}
    \begin{tabular}{l c c c} 
        \hline
        Filter & $\lambda_0$ [$\mu$m] & $F_{\nu}^{\rm 0917}$ [$\mu$Jy] & \added{$F_{\nu}^{\rm SW}$ [$\mu$Jy] } \\
        \hline \hline
        HSC\_$g$ & 0.48 & <0.08 & \added{0.5$\pm$0.06}$^\dagger$ \\
        HSC\_$r$ & 0.62 & <0.12 & \added{0.69$\pm$0.07}$^\dagger$ \\
        HSC\_$i$ & 0.77 & <0.14 & \added{0.66$\pm$ 0.07}$^\dagger$\\
        HSC\_$z$ & 0.89 & <0.23 & \added{0.65$\pm$0.07}$^\dagger$ \\
        VIKING\_$z$ & 0.9 & <0.7 & \added{<0.7} \\
        HSC\_$y$ & 0.97 & <0.32 & \added{0.7$\pm$0.1}$^\dagger$ \\
        VIKING\_$Y$ & 1.0 & <1.5 & \added{<1.5} \\
        VIKING\_$J$ & 1.25 & <1.8 & \added{<1.8} \\
        VIKING\_$H$ & 1.6 & <3.4 & \added{<3.4} \\
        HAWKI\_\ks & 2.2 & 3.1$\pm$0.1 & \added{1.2$\pm$0.1}\\
     \hline
    \end{tabular}
     \begin{tablenotes}
     \item \added{[$\dagger$]{{\footnotesize 10\% uncertainties have been added in quadrature to account for absolute calibration uncertainties.}}}
     \end{tablenotes}
     \end{threeparttable}
    \label{tab:optical_flux}
\end{table}

\section{Results}
\label{sec:results}

\added{As the molecular lines we were targetting are not detected, we} focus \deleted{here} on the broad-band SEDs in the optical to near-IR and radio to mid-IR\deleted{frequencies}, and their implications in terms of luminosity and radio loudness. \deleted{We further discuss the non-detection of the molecular lines in Section~\ref{sec:z_0917}.}

\subsection{Optical to near-IR SED}
\label{sec:optical_SED}

\subsubsection{\added{South-western source}}
\begin{figure}
    \centering
    \includegraphics[width=0.5\textwidth,trim=0 0 0 0,clip]{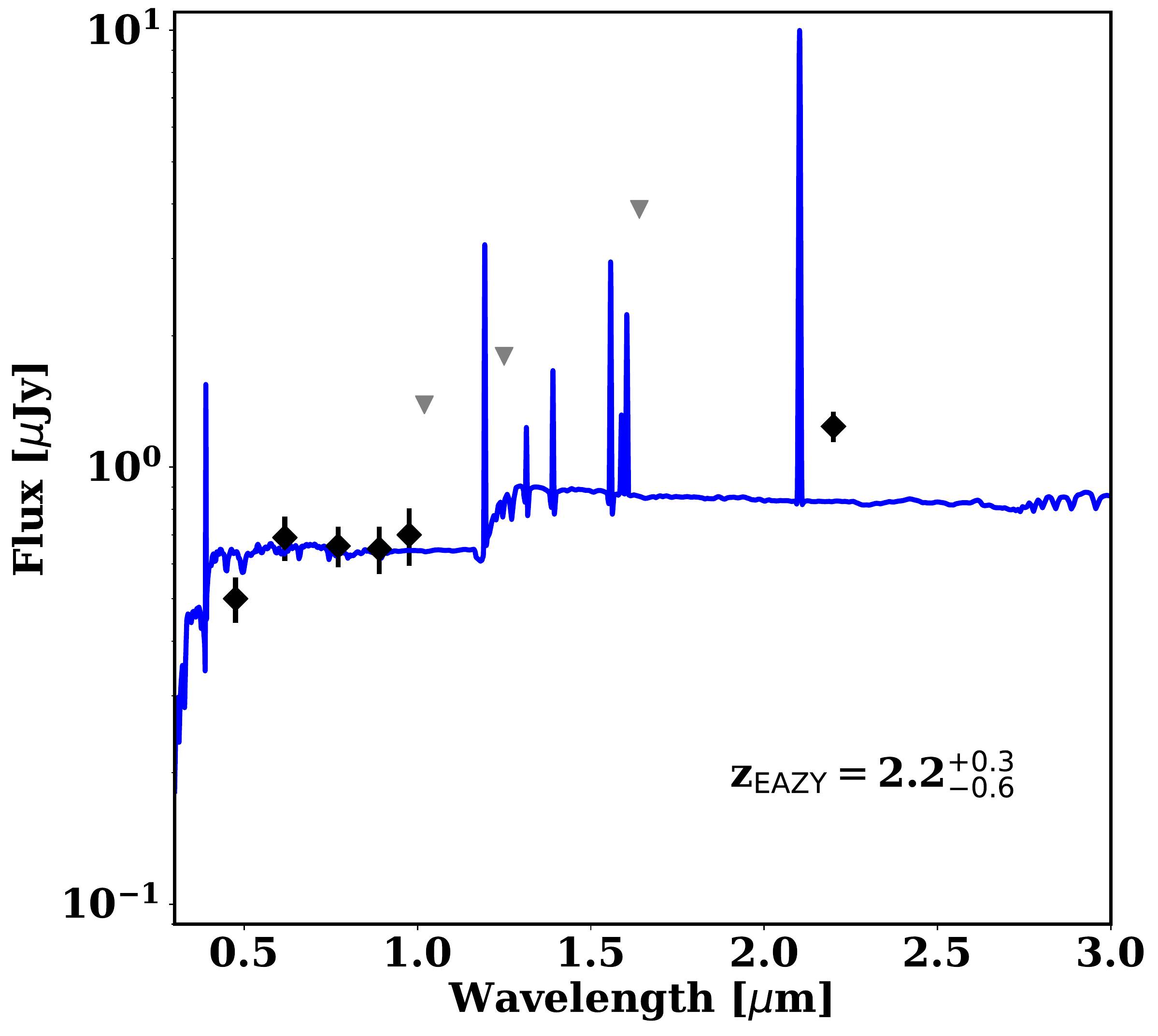}
    \caption{\added{
Optical to near-IR SED for the south-western source identified in Fig.~\ref{fig:allradio}. Flux densities are reported in Table~\ref{tab:optical_flux}. The black diamonds are the detections and the grey downward triangles the 3$\sigma$ upper limits. The template overlaid is the best fit from the {\sc EAZY} fitting at the redshift indicated with the 68th-percentile reported as the uncertainties (see Section~\ref{sec:optical_SED}).}}
    \label{fig:SW-source}
\end{figure}

\added{
Fig.~\ref{fig:SW-source} presents the optical to near-IR SED of the source south-west of the host in Fig.~\ref{fig:allradio}. Aperture photometry is applied in the same fashion as described in D20 with a 0.8-arcsec aperture. The flux densities are extracted at the coordinates in the HAWKI image and reported in Table~\ref{tab:optical_flux}. The source is detected in the HSC images, but undetected in VIKING. Using {\sc EAZY} \citep[a photometric redshift code; ][]{brammer_eazy:_2008} on this photometry with all parameters at their default value, leads to a redshift estimate of $z_{\rm phot}=2.2^{+0.3}_{-0.6}$ taking the 68th-percentile confidence interval from the resulting redshift distribution. The best-fitting template is also shown in Fig~\ref{fig:SW-source}.}

\subsubsection{\added{Host galaxy, \oddball}}

\begin{figure*}[t]
    \centering
    \includegraphics[width=0.9\textwidth, trim= 0 0 0 0,clip]{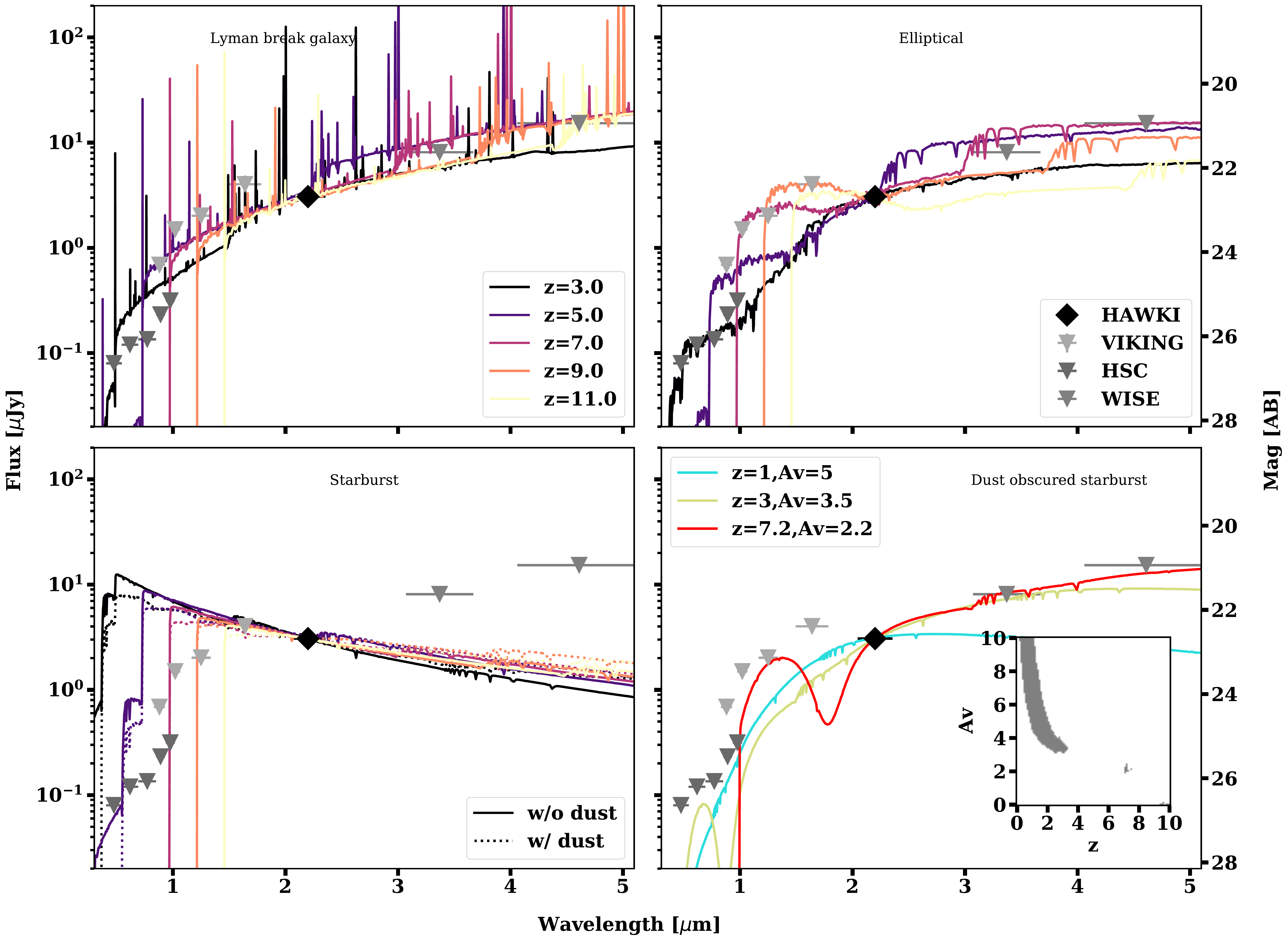}
    \caption{Optical to near-IR SED of \oddball\ with each panel overlaying different galaxy templates over a range of redshifts (each template is normalised to the \ks-band detection). The diamond indicates the \ks-band flux density from D20 \added{(note that the uncertainty is smaller than the symbol)}, and the downward pointing triangles are the 3$\sigma$ upper limits from the VIKING and HSC images using the same aperture as that used to measure the \ks-band flux density. The grey shaded area in the fourth panel inset indicates the permissible solutions for the extinction \deleted{in respect to the upper limits} \added{(note the small island of possible solutions at $z\sim7$ and $A_{\rm V}\sim$2)}. See Section~\ref{sec:optical_SED} for more details about the templates and Section~\ref{sec:money_plot_near-IR} for a discussion.}
    \label{fig:optical_SED}
\end{figure*}

In Fig.~\ref{fig:optical_SED} we present the optical to near-IR SED of \oddball. Despite the new deep optical photometry from HSC, our HAWKI \ks-band observation (see D20) provides the only detection. There is a break of \added{more than} an order of magnitude between the $K_{\rm s}-$band detection and the HSC $i$-band upper limit. If this is due to the Lyman break, then the galaxy must lie at $z \gtrsim 6.5$, or alternatively the break suggests \added{a very red colour due to} significant dust obscuration. To further investigate this potential redshift constraint, we overlay four different templates in Fig.~\ref{fig:optical_SED} \added{\citep[three of which are taken from \pegase, a galaxy evolutionary code predicting SEDs for a given scenario of evolution;][]{fioc_pegase3_2019} at a range of redshifts. These templates are as follows.} 
\begin{enumerate}
    \item \added{A Lyman-break galaxy (LBG) from \cite{alvarez-marquez_rest-frame_2019}.}
    \item \added{A \pegase~elliptical (E) galaxy template described in \citet{drouart_disentangling_2016}, assuming the maximum age hypothesis as per \cite{seymour_massive_2007} and $z_{\rm form}=20$.}\footnote{This change from the original $z_{\rm form}=10$ introduces a small shift in the age of the corresponding colours at $z=0$. The age shift corresponds to the difference in look-back time between $z=10$ and $z=20$, $\sim$300\,Myr earlier under the cosmology adopted here, but is necessary to explore the full redshift range.}
    \item \added{A \pegase~starburst (SB) template, with and without internal dust extinction, corresponding to a single stellar population with an age of 10\,Myr (the amount of dust extinction is calculated consistently within the code assuming a spherical geometry).}
    %; see \citet{fioc_pegase3_2019} for more details)}
    \item \added{The same \pegase~SB template as above but with additional dust extinction (in the form of a dust screen) making use of the \citet{fitzpatrick_correcting_1999} law.} 
\end{enumerate}

%\deleted{The first template is a Lyman-break galaxy (LBG) from \cite{alvarez-marquez_rest-frame_2019}. The second is the \pegase\,elliptical galaxy template used in \citet{drouart_disentangling_2016}, assuming the maximum age hypothesis as per \cite{seymour_massive_2007}. The third is a starburst (SB) template, with and without internal dust extinction from \pegase~ (...)}
%\deleted{\citep[a galaxy evolutionary code predicting SEDs for a given scenario of evolution, ][]{fioc_pegase3_2019}
%\deleted{, corresponding to a single stellar population (SSP) with an age of 10\,Myr (the amount of dust extinction is calculated consistently within the code assuming a spherical geometry; see \citet{fioc_pegase3_2019} for more details). Finally, we apply additional dust extinction on the SB template making use of the \citet{fitzpatrick_correcting_1999} law as the fourth template. The inset shows the permissible solution given our upper limits.}

\added{We normalise each of the four templates to the detection in \ks-band 
%\deleted{For each of these four templates, we normalise to the detection in \ks-band,} 
and apply an additional extinction shortward of the rest-frame Ly$\alpha$. We use the formalism of \cite{fan_observational_2006} to characterise this extinction due to neutral hydrogen (HI) in the intergalactic medium (IGM)}. %\deleted{intergalactic medium (IGM) extinction due to the neutral hydrogen gas remaining from reionisation following the formalism from \cite{fan_observational_2006}}. 
In brief, this absorption starts to be significant at $z>4$, and reaches full effect at $z\sim6$ due to the evolution of the reionisation of the IGM. At \added{$z>6$} \deleted{highest redshifts}, essentially all photons shortward of Ly$\alpha$ are absorbed by intervening neutral HI gas. 

% new stuff
\added{We can see that some templates are acceptable fits and some are not.} 
\begin{enumerate}
    \item \added{The LBG template is consistent with the data for $z\gtrsim 7$ (lower-redshift options exceed the HSC fluxes).}
    \item \added{The elliptical template does not work for any redshift as it cannot reproduce the strong break.}
    \item \added{The starburst template is too blue to fit the constraints even with some internal dust. }
    \item \added{For the starburst template with a dust screen, the acceptable parameter space of redshift and $A_{\rm V}$ is more complex; the inset shows the permissible solutions (given the 3$\sigma$ upper limits). }
\end{enumerate}

\added{The two possible highly dust-obscured starburst solutions are (i) $A_{\rm V}>3.5$ with a lower-redshift solution ($z<3$), and (ii) $A_{\rm V}\sim2$ with a high-redshift solution ($z\sim7$).}

%\deleted{Qualitatively, if the source is not heavily dust obscured, the $z=5-7$ \added{LBG, elliptical and SB} templates suggest that we should have detected \oddball\, in the HSC \deleted{and/or}, VIKING $zYJH$ \added{or WISE} data. \added{The first three panels of Fig.~\ref{fig:optical_SED} therefore suggests that \oddball\ is located at $z>7$.} In the case of a \added{highly} dust-obscured source, favoured scenarios \replaced{are (i) $A_{\rm V}>4$ with a lower redshift solution ($z<2$), (ii) $A_{\rm V}\sim3.5$ with an intermediate redshift solution ($2<z<3$) or (iii) $A_{\rm V}\sim2$ with at a high redshift solution ($z\sim7$) }{are $A_{\rm V}=4$ and a lower redshift solution ($2<z<6$) or $A_{\rm V}>5$ and a lower redshift solution ($z<2$)are $A_{\rm V}=4$ and a lower redshift solution ($2<z<6$) or $A_{\rm V}>5$ and a lower redshift solution ($z<2$)}. }

Again, the HSC data are \deleted{here} the strongest constraints for the lower limits on the obscuration and redshift, but the {\it WISE} data provide us with the strongest constraints for the upper limits on the obscuration and redshift. \deleted{We argue against the lowest-redshift solutions ($z<2$) and more heavily obscured solutions.} \added{An $A_{\rm V}>3$ solution corresponds to an extreme obscuration value even when compared to infrared luminous galaxies, i.e. containing a large amount of dust \citep[e.g.][for {\it Herschel}-selected galaxies]{buat_goods-herschel_2011}, which typically present $A_{\rm V}\sim1$. Larger values of $A_{\rm V}$ are usually associated with AGN \citep[e.g.][]{drouart_jet_2012}. Given the highest $A_{\rm V}>3$ solutions are associated with the low-redshift solutions ($z<3$), more obscuration translates into larger far-IR flux densities, which would be detected with {\it Herschel}. Therefore, the {\it Herschel} upper limits for \oddball\ gives little support to these low-redshift solutions.} 
%\deleted{Any larger $A_{\rm V}$ would imply larger dust masses: $M_{\rm dust}>10^{7}$\msun\ assuming a 4\,kpc radius \citep{ferrara_atlas_1999}, the most mass conservative $A_{\rm V}$-$M_{\rm dust}$ conversion. However, this value is in direct tension with the dust mass allowed from the \mrmoose\ fit (see Section~\ref{sec:radio_SED}). In conclusion, while a moderately dust-obscured source ($A_{\rm V}v\sim4$) at an intermediate redshift ($2<z<6$) cannot be excluded, there is very little room in the dust extinction parameter space to be consistent with our data, which decrease the likelihood of this possibility.}

\subsection{Radio to mid-IR SED}
\label{sec:radio_SED}

\begin{figure*}
  \centering   
  \begin{overpic}[width=1.0\textwidth]{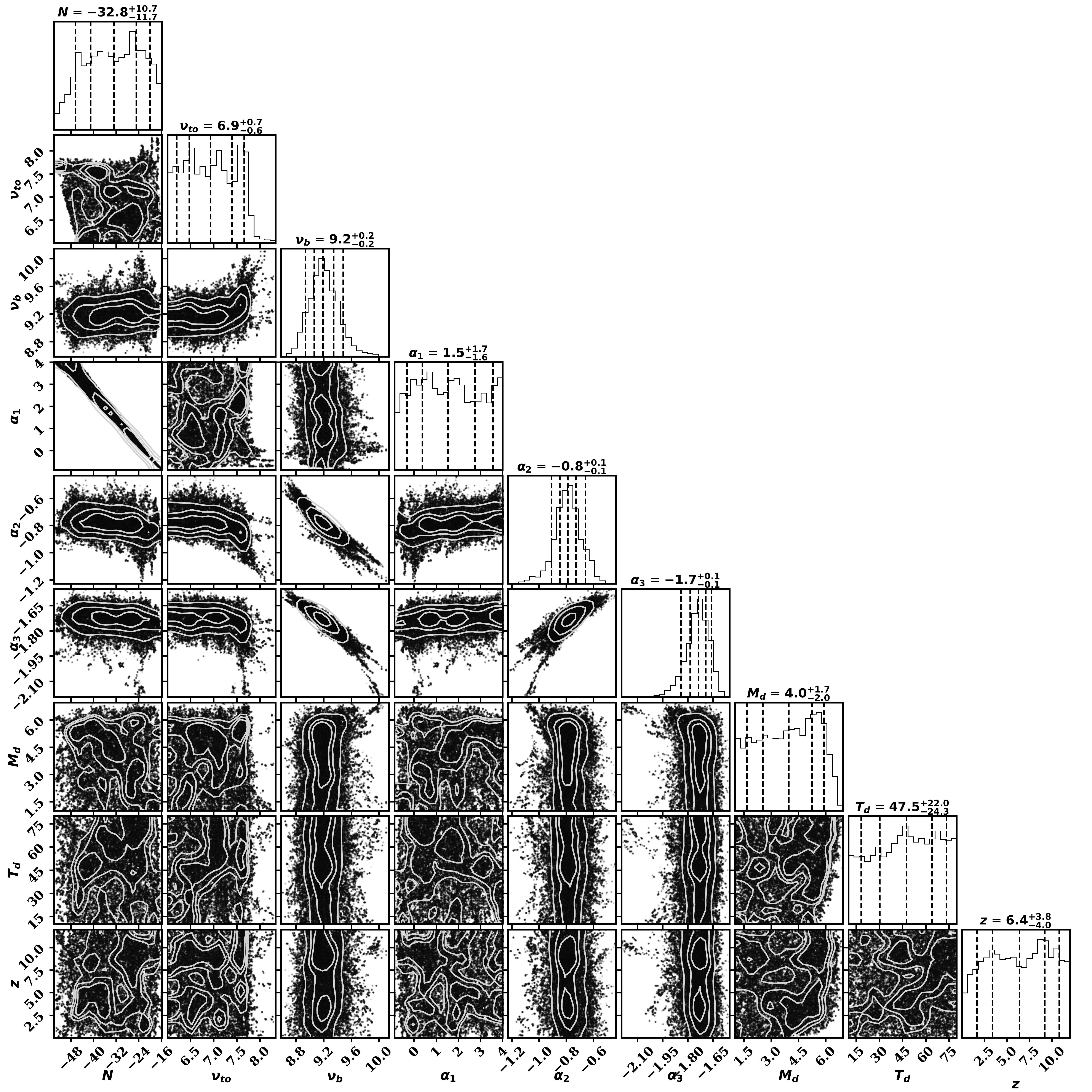}
     \put(39,71){\includegraphics[width=0.65\textwidth,trim={0.55cm 0.3cm 0 0},clip]{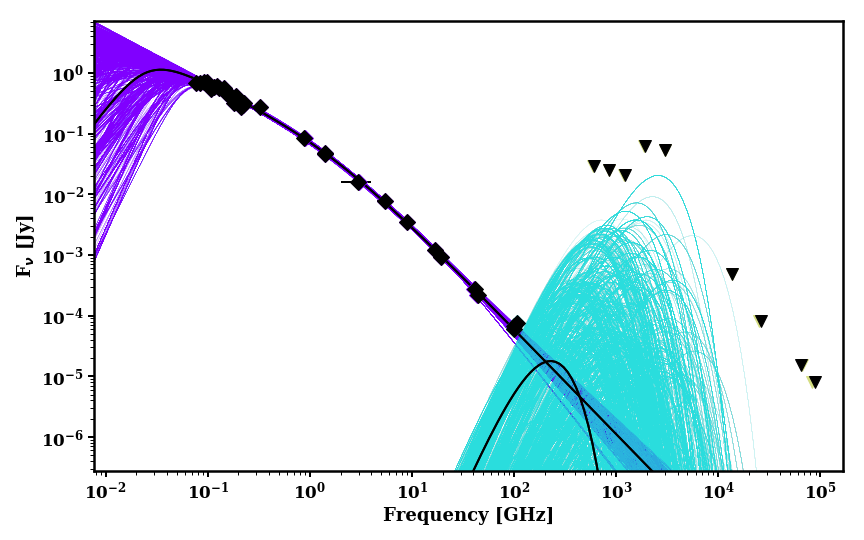}}  
  \end{overpic}
    \caption{{\it Upper right:}  Radio to far-IR SED of available data for \oddball\, fitted with \mrmoose. Diamonds and triangles are detections and 3$\sigma$ upper limits respectively. The two solid black lines are the best fits for each component (the triple power law and modified blackbody; see Eq.~\ref{eq:mbb}), with the purple and blue lines showing the probability distribution for each component respectively. {\it Lower left:} `Corner plot' of the marginalised probability density distributions for all parameters with \replaced{the median}{best-fit} value and uncertainty \added{as the interquartile range} at the top of each column. \added{The vertical lines are the 10, 25, 50, 75 and 90th percentiles, respectively}. We report the units and best constraints for the parameters in Table~\ref{tab:sed_results}.} %\warning{update the figure with new WISE limits}.}
    \label{fig:mrmoose}
\end{figure*}

We present the radio to mid-IR SED in the upper-right of Fig.~\ref{fig:mrmoose}. The new VLA data points confirm the \added{spectral} break \added{seen} at GHz frequencies \deleted{, as presented} in D20. Also, the three ALMA data points potentially suggest a change of slope (see Table~\ref{tab:radio_flux}), indicating a possible radio core component or a possible dust contribution. Even though the first option cannot be completely excluded, it seems unlikely given that the \replaced{D20}{steep-spectrum} selection tends to favour sources aligned in the plane of the sky (type 2 AGN), where the core contribution is minimal due to the lack of Doppler boosting \citep[][]{drouart_jet_2012}.

While meaningful constraints on dust properties are not possible at this stage given the lack of a detection in the mid-IR, we can use \mrmoose~ \citep[\added{an advanced Bayesian multi-component fitting code treating consistently the upper limits;}][]{drouart_mrmoose:_2018} to determine an upper limit to the contribution from dust at 100\,GHz, and predict at which frequency this contribution would become dominant.

We have added our new VLA and ALMA continuum points, radio data from the literature, and the {\it WISE} and {\it Herschel} infrared upper limits to the SED \added{(see Table~\ref{tab:radio_flux})}. In addition to fitting the triple power law (see D20; Eq.~4), we include a new component of the form of a modified blackbody, calculated as follows:
\begin{equation}
    f_\nu=(1+z)\frac{M_{\rm d}}{D_L^2}\kappa_{\rm abs}(\nu_0)\left(\frac{\nu}{\nu_0}\right)^{\beta}B_\nu(T_d),
    \label{eq:mbb}
\end{equation}
with $z$ the redshift, $D_L$ the luminosity distance, $\kappa_{abs}$ the grain absorption cross section per unit mass (also referred to as the dust emissivity), $\beta$ the power law index for the dust emissivity, $\nu_0$ the reference frequency, $M_d$ the mass of the dust, $B_\nu$ the classical blackbody function (Planck's law), and $T_d$ the temperature of the dust. For the sake of simplicity, we assume\footnote{\added{We also ran \mrmoose\ with the $\beta$ parameter free in the 1$<\beta<$3.5 range, resulting in a factor $\sim$2 change in the dust mass results.}} that $\beta$=2.08, $\nu_0=$250\mum\ and $\kappa_0$(250\mum)=4.0 cm$^2$g$^{-1}$ \citep{draine_infrared_2007}. We note that plausible different assumptions for $\beta$ and $\kappa_{abs}$ would introduce a factor two to four change in the dust mass, but would hinder any direct comparison with other samples \citep[for an extensive explanation, see][]{bianchi_vindicating_2013}. This leaves us with three free parameters \added{for the dust component}: $T_d$, $z$ and $M_d$. 

\added{We simultaneously fitted the triple power law and the dust component to the radio to mid-IR data with a set of uniform priors to the nine free parameters (six for the triple power law and three for the dust component).} The results of the fitting are overlaid on the SED in the upper right of Fig.~\ref{fig:mrmoose} and the parameter constraints in the corner plot in the lower left. The best-fit parameters are presented in Table~\ref{tab:sed_results}, \added{along with the range of the uniform considered priors}. As one would expect, \deleted{we clearly see that} neither the redshift nor the dust temperature\footnote{Note that the rising floor temperature imposed by the cosmic microwave background (CMB) with increasing redshift is not taken into account in the priors.} are constrained. However, the dust mass limit is particularly interesting\added{: the 90th percentile for the dust mass distribution gives $M_{\rm dust}=10^{5.7}$\msun. Accounting for unknown and large systematic uncertainties, a conservative upper limit of }\deleted{no solution with} $M_{\rm dust}<10^7$\msun\ is acceptable given our data, and this result is independent of both the redshift and dust temperature. 
 
While the maximum dust contribution at 100\,GHz appears to be $\sim10$\% \deleted{at maximum}, only \deleted{clear} detections at $>200$\ghz~with \deleted{a} higher resolution would definitively settle the question of the presence of an upturn in the SED and its origin. A \added{radio} core would stay unresolved and would \deleted{likely} present a flat\deleted{ter} spectral index, while dust emission would start to be resolved, show no axial symmetry and \added{the Rayleigh--Jeans tail of cold dust} \deleted{slope} would \replaced{result in}{present} a \deleted{more} positive slope. \deleted{Moreover, the radio spectral slope provides key information regarding the nature of this emission: AGN radio cores tend to be flat spectrum, while the Rayleigh--Jeans tail of cold dust produces a steep spectrum.}

\added{
Finally, a conversion between the obscuration (from Section~\ref{sec:optical_SED}) and the dust mass leads to interesting limits. Assuming (i) a conservative 4-kpc radius for our dust emitting region (equivalent to $<0.5$\,arcsec at $z>1$, similar to the limits from our \ks-band and ALMA observations; see Section~\ref{sec:size}), (ii) a $M_{\rm dust}<10^{7}$\msun\ limit from the \mrmoose\ fitting and, (iii) the most mass conservative $A_{\rm V}$-$M_{\rm dust}$ conversion from \citep[][ see the constant sphere approximation in the appendix]{ferrara_atlas_1999}, we obtain a upper limit of $A_{\rm V}\lesssim 4$. Any larger physical size, lower dust mass, or different geometries would lead to a lower $A_{\rm V}$. This constraint has further implications for the possible redshift of \oddball, which are discussed in Section~\ref{sec:z_0917}.}
%any large $A_{\rm V}$ is in direct tension with the estimated dust mass limit. 

%Any larger $A_{\rm V}$ would imply larger dust masses: $M_{\rm dust}>10^{7}$\msun\ assuming a 4\,kpc radius \citep{ferrara_atlas_1999}, the most mass conservative $A_{\rm V}$-$M_{\rm dust}$ conversion. However, this value is in direct tension with the dust mass allowed from the \mrmoose\ fit (see Section~\ref{sec:radio_SED}). In conclusion, while a moderately dust-obscured source ($A_{\rm V}v\sim4$) at an intermediate redshift ($2<z<6$) cannot be excluded, there is very little room in the dust extinction parameter space to be consistent with our data, which decrease the likelihood of this possibility.

% \begin{figure*}
%   \centering   
%   \begin{overpic}[width=1.0\textwidth]{0917_pub_IRradio_SED_w200_s8000_triangle.pdf}
%      \put(44,68){\includegraphics[width=0.6\textwidth,trim={0.55cm 0.3cm 0 0},clip]{0917_pub_IRradio_SED_w200_s8000_SED_fnu_spag.pdf}}  
%   \end{overpic}
%     \caption{{\it Upper right:} complete radio-far-IR SED of available data for \oddball fitted with \mrmoose. Triangles and diamonds are upper limits and detections respectively. Purple is a triple power law with two break frequencies and blue is a modified black body (see Eq.~\ref{eq:mbb}). {Lower left:} marginalised probability density distribution for the parameters. \warning{not the final version}}
%     \label{fig:mrmoose}
% \end{figure*}

\begin{table}[t]
    \caption{Results from the observed-frame radio to mid-IR SED fitting. We refer the reader to D20 for the triple power law equation and description of the fitting procedure, as well as Eq.~\ref{eq:mbb} for the modified blackbody. We report the 25th--75th percentiles as uncertainties.}
    \centering
    \begin{threeparttable}
    \begin{tabular}{c c c}
        \hline
        Parameter & \added{Uniform prior} & \added{Value} \\
        & \added{range} & \\
        \hline \hline
        $N$ & $-$55, $-5$ & $-32.8^{+10.7}_{-11.7}$ \\
        $\log~\nu_{\rm b\_to}^\dagger$ & 6, 8.5 & 6.9$^{+0.7}_{-0.6}$ \\
        $\log~\nu_{\rm b\_b}$  & 8.5, 12 & 9.2$^{+0.2}_{-0.2}$ \\
        $\alpha_{1}^\dagger$ & $-1$, 4 & 1.5$^{+1.7}_{-1.6}$ \\
        $\alpha_{2}$ & $-3$, 0 & -0.8$^{+0.1}_{-0.1}$ \\
        $\alpha_{3}$ & $-4$, $-1$ & $-1.7^{+0.1}_{-0.1}$ \\
        $\log~M_{\rm d}$ [\msun] & 1, 10 & 4.0$^{+1.7}_{-2.0}$ \\
        $T_{\rm d}^\dagger [K]$ & 10, 80 & 47.5$^{+22.0}_{-24.3}$ \\
        $z^\dagger$ & 0, 12 & 6.4$^{+3.8}_{-4.0}$ \\
        \hline
    \end{tabular}
     \begin{tablenotes}
     \item [$\dagger$]{{\footnotesize Considered not constrained by the fit.}}
     \end{tablenotes}
    \end{threeparttable}
    \label{tab:sed_results}
    \end{table}

\subsection{An extreme radio-to-near-IR ratio}
\label{section:flux_ratios}

% \warning{addrefs}: 
% Kellermann, Pauliny-Toth \& Williams 1969 (3C); 
% Allington-Smith 1982 (6C); 
% Lilly \& Longair 1984 (3C); 
% Eales et al. 1985 (6C); 

\begin{figure*}[t]
    \centering
    \includegraphics[width=0.75\textwidth]{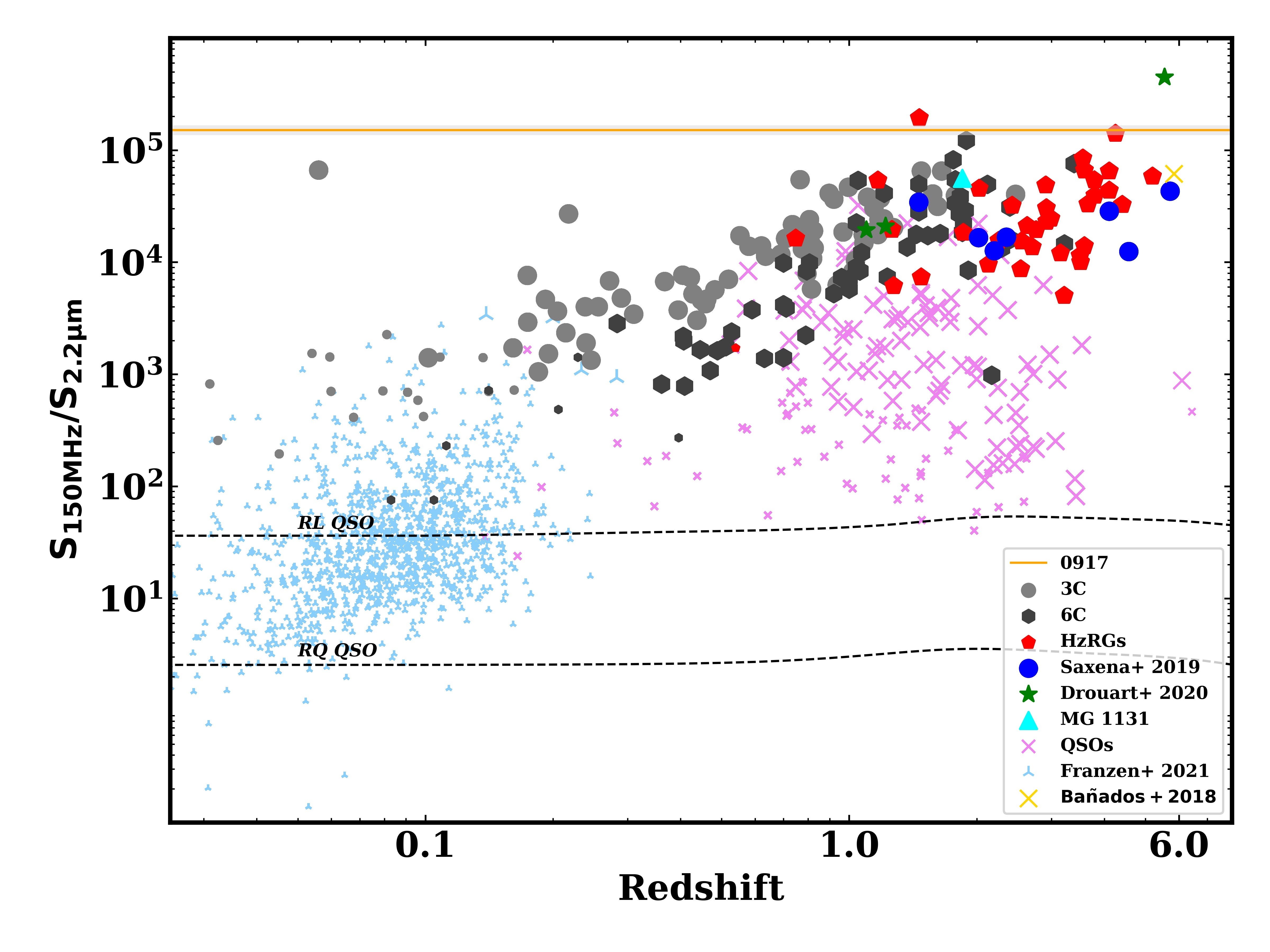}
    \caption{The observed-frame $150\,$MHz to \ks-band flux density ratio as a function of redshift for \added{different classes of radio-powerful AGN} \deleted{radio sources of different types} (see legend). Larger symbols are those with a luminosity $L_{\rm 150MHz}>10^{27}\,$WHz$^{-1}$. The two template tracks are from \citet{elvis_atlas_1994} \added{for a radio-loud and radio-quiet QSO}. The orange horizontal line at a value of $10^{5.1}$ \replaced{represents \oddball. Note that uncertainties are reported in grey, but are similar to the width of the line.}{with light grey shading denoting the uncertainty, represents \oddball}. Further details can be found in Section~\ref{section:flux_ratios}.}
    \label{fig:Krad}
\end{figure*}

Part of our selection technique (D20) for very high-redshift radio galaxies is their brightness at low-frequencies ($S_{\rm 150\,MHz}>100\,$mJy) and faintness in \ks-band (\ks$>21.2$). This technique has been used by several groups before us \citep[e.g.][]{de_breuck_optical_2002} and is designed to select galaxies with powerful jets observed in the radio, but at high redshift \added{where}\deleted{as} the host galaxy light is faint. To investigate where \oddball\, lies with respect to the radio-\replaced{powerful}{loud} AGN population, we present in Fig.~\ref{fig:Krad} the observed-frame 150\mhz\,to 2.2\mum\,(the latter corresponding to $K$- and \ks-band) ratio \added{plotted against redshift} for a selection of radio sources from the literature. 

As the redshift of \oddball\, is unknown, we mark it by a horizontal line. At high redshift we include the other three sources from D20, the sample from \citet[][]{saxena_nature_2019} and the radio-loud QSO from \citet{banados_800-million-solar-mass_2018}. The two radio-loud QSOs at $z>6$ come from \citet{ighina_radio_2021} and \citet{spingola_parsec-scale_2020}. MG~1131+0456 is a radio-loud lensed galaxy \citep{stern_redshift_2020}. Furthermore, we used data from the following studies: 3C sample \citep[][]{lilly_stellar_1984}, 6C sample \citep[][]{eales_sample_1985}, compilation of known HzRGs at $z\ge 2$ \citep[][]{miley_distant_2008}, QSOs cross-matched with TGSS \citep[][]{paris_sloan_2018}, and a complete sample of local, GLEAM-selected radio-loud AGN from a cross-match with 6dFGS, (Franzen et al., \replaced{submitted}{in prep}).

%We also made use of data available in the NASA/IPAC Extragalactic Database (NED)\footnote{https://ned.ipac.caltech.edu/}.\warning{should we really add the individual references if there aren't too many?} Where $150$-MHz flux densities were not available in the literature, we extrapolated from higher-frequency data using known two-point spectral indices.\warning{how often was this the case?}

\deleted{We can see that,} Regardless of redshift, \oddball\, has the third-most extreme flux density ratio, with our previous $z=5.55$ discovery (GLEAM\,J0856+0224 from D20) being the most extreme. Barring Cygnus A, the high-ratio datum at $z<0.1$, we observe a general trend of increasing ratio with redshift. This result is not a selection effect of the populations used (the low-redshift 3C and Franzen et al. samples are both complete in brightness),  but rather likely driven by the proliferation of powerful jets earlier in the Universe when the black holes in galaxies were more active. 

Based on rest-frame near-IR data of a large sample of HzRGs, \citet{seymour_massive_2007} showed such galaxies consistently have stellar masses of $10^{11} - 10^{11.5}$M$_\odot$ even out to the highest redshifts probed in that work ($z\sim 5$).  We also note the differing $k$-corrections for the radio and infrared data.  \replaced{While both wavelength ranges will sample a falling portion of the SED, they will change at different rates depending on (i) the steepness of the radio spectra ($\alpha<-0.7$, and decreasing with increasing redshift at higher rest-frame frequencies) and (ii) the amount of star formation and dust extinction in the observed-frame 2.2-\mum~ emission (at an increasingly shorter wavelength with increasing redshift).}{while the steep radio spectra ($\alpha < 0$) will sample a rising portion of the SED at higher redshifts, observed $2.2$-$\mu$m emission will sample a falling portion of the stellar population SED at higher redshifts, particularly if dust absorption is significant.}

%By and large the stellar masses of the host galaxies are comparable at high and low-redshift \citep{seymour_massive_2007}. There is also the issue of k-correction. At increasing redshift the radio samples lower frequency emission which is often increases due to steep spectra ($\alpha<0$) whereas the \ks-band samples shorter wavelengths where often the emission decreases due to absorption by dust or old stellar populations. 

\deleted{Interestingly, the ratio for \oddball\, is similar to that for Cygnus A at $z=0.056$. Cygnus A is a unique source in the local Universe with no similar sources until at least $z=0.2$. While it has a flux ratio comparable to many of the most powerful radio galaxies at $z>1$, it has a 2.2-\mum\ flux density at least three orders of magnitude brighter than \oddball. \replaced{Curiously}{Interestingly}, if one takes the SED of Cygnus A and shifts it to higher redshifts, this observed flux density ratio increases to $\ge 10^6$ by $z=4$.}

%However, the faintness of \oddball in $K$-band suggests it is at high redshift.

%\warning{also include Liu et al. 2020 RL high-z QSO results, and z=6.44 QSO? Plus Belladitta blazar?}. 

%\section{Constraining the redshift of \oddball}
\section{\added{Analysis of observable quantities}}
\label{sec:z_0917}

As none of the expected emission lines were detected in either the deeper ALMA or the VLA observations, the D20 detections must have been noise peaks regrettably aligned in frequency. The logical conclusion is therefore that \oddball\, is not at $z=10.15$, and that the D20 detections were instead a manifestation of the effect described in the appendix of \cite{krips_co_2012}, where the widening bandwidth in modern interferometers increases the potential for spurious detections at low signal-to-noise.

 However, such a large amount of data over the electromagnetic spectrum provides us with numerous upper limits, and therefore information which can be used to make an \deleted{educated} estimate for the redshift of \oddball. Fig.~\ref{fig:money_plot} synthesises all of our \added{constraints from the} data \added{presented in Section~\ref{sec:data}}. For each subplot ($a$ to $e$), we investigate the following observable quantities and our constraints on them across $0.01<z<12$. 
 
 \begin{enumerate}[label=\alph*)]
     \item The near-IR flux density. 
     \item \added{The} radio and sub-mm flux densities. 
     \item The angular size of the host galaxy and radio emission. 
     \item The luminosity of the CO gas, and
     \item Whether other molecular lines \deleted{could be detectable} \added{fall within the observing windows of our ALMA and VLA observations}.
 \end{enumerate}

\added{We investigate the constraints of each of the above observables separately in this section, for clarity, due to the assumptions which go into each part. We keep the discussion of the joint constraints for Section~\ref{sec:disc}.} For comparison to \oddball, in each subplot we also present the corresponding properties \deleted{(and the references)} of {Cygnus A \citep[$z=0.0561$; ][]{mazzarella_molecular_1993,condon_nrao_1998,skrutskie_two_2006},} the Spiderweb galaxy \citep[$z=2.16$; ][]{emonts_co1-0_2014,seymour_massive_2007,de_breuck_spitzer_2010} and GLEAM\,J0856+0224 ($z=5.55$; D20), in order to quantitatively compare \oddball\, with known sources of the same class.

\subsection{Near-IR properties}
\label{sec:money_plot_near-IR}

Fig.~\ref{fig:money_plot}a presents the predicted $K_s$-band \added{flux density } of a stellar population as a function of redshift based on an assumed evolutionary history. These calculations \added{are} made using \pegase, \added{and assume the maximum age approximation \citep{seymour_massive_2007}.} \deleted{, provide insight in removing unrealistic redshift solutions} Given that \oddball\ is \deleted{only} detected in a single near-IR band, we restrict this analysis to \replaced{four modelled}{three} templates, \added{some of which are presented in Section~\ref{sec:optical_SED}}:  an elliptical (E), a spiral (S), a 10-Myr\ starburst\footnote{We checked the effect of changing the starburst age in the 1--20\,Myr range (translating observationally to a mass-to-luminosity ratio change). It translates to an increased scatter of one order of magnitude below/above the reported 10-Myr template.} (SB) and an obscured 10-Myr starburst (with $A_{\rm V}=4$)\footnote{We note that \oddball\, is unlikely to be a low-luminosity hot dust-obscured galaxy \citep[HotDOG, i.e., a galaxy with mid-IR emission dominated by an obscured torus;][]{eisenhardt_first_2012,tsai_most_2015} as the spectral slope in the 2.2-3.6\mum range is $\alpha>-2$, flatter than the $\alpha\approx -3$ found in the short-wavelength regime of HotDOG galaxy SEDs \citep{fan_infrared_2016}.} \added{in order to normalise the track to a given stellar mass and assess the impact of star formation histories on the \ks-band flux density with respect to redshift.}

\deleted{We assume a galaxy formation redshift of $z_{\rm form}=20$ for the elliptical and spiral templates. This choice is different from the  $z_{\rm form}=10$ from \pegase2 \mbox{\citep{fioc_pegase:_1997}} in order to push the formation of the galaxy to a redshift \added{above} \deleted{higher than} our considered range and avoid effects related to initial conditions. This change introduces a small shift in the age of the corresponding colours at $z=0$. The age shift corresponds to the difference in look-back time between $z=10$ and $z=20$, $\sim$300\,Myr earlier under the cosmology adopted here. While the effect is minimal at low redshift, one should be careful when using these templates in the highest-redshift regime. }

\deleted{The SB template is derived from a single collapse of gas with no further star formation. We set the age to 10\,Myr and redshift this template accordingly. We take this approach in order to explore the redshift effect, leading to a change in the \ks-band flux, as we probe further into the UV rest-frame with increasing redshift. We also checked the effect on the flux of changing the starburst age in the 1-20\,Myr range (translating observationally to a the mass-to-luminosity ratio change). In Fig.~\ref{fig:money_plot}a this corresponds to an increased scatter of one order of magnitude below/above the reported 10\,Myr template.}

We show the tracks for a 10$^{12}$\msun\, galaxy for all \replaced{four}{three} scenarios. Bright atomic line contributions are not included in these templates, which could affect certain redshift windows of the spiral and SB templates (shaded red regions in Fig.~\ref{fig:money_plot}a). Strong lines would result a slightly overestimated stellar mass in these redshift ranges. We note that any AGN contribution to the \ks-band flux would only decrease the stellar mass estimates. \deleted{This is however unlikely due to the steep-spectrum selection picking preferentially picking galaxies with radio jet axis in the plane of the sky \mbox{\citep{drouart_jet_2012}}, as well as the \ks-band magnitude, which is higher than any typical quasar at high redshift \mbox{\citep[e.g.][]{banados_pan-starrs1_2016}}.}

The upper limits in the optical (Section~\ref{sec:optical_SED}) are not used in these constraints, but we checked the relative depth of each band and their relative constraints. The \ks-band data provide the best constraints due to being the longest-wavelength \added{near-IR} observation available and comparatively the deepest image. \deleted{However, the optical upper limits seem to favour $z>5$ (e.g. Fig.~\ref{fig:optical_SED})}. We note several implications of the single broadband detection as follows.
\begin{itemize}
    \item For $z>2$, the system is massive, \mstel>10$^{11}$\msun, assuming 100\% stellar light and the elliptical template. The spiral template is implausible at $z>4$, as it would require a system with \mstel$>10^{12}$\,\msun. The 10-Myr-old SB template reproduces the \ks-band emission for \mstel$>10^{10}$\,\msun\, at $z>4$. Any dust obscuration requires an even greater stellar mass. 
    \item For $z<1$, the system would likely have a low galaxy mass, \mstel $<10^{10}$\,\msun, and would therefore likely be classified as a dwarf galaxy.
    \item The \replaced{$z>7$}{$z>5$} solution is favoured from the optical upper limits; see Fig.~\ref{fig:optical_SED} and Section~\ref{sec:optical_SED}.
\end{itemize}
 
\subsection{Radio and sub-mm properties}
\label{sec:money_plot_radio}

\deleted{Our new observations and literature data add eight data points in this frequency range as reported in Table~\ref{tab:radio_flux}, and confirm the curvature observed in the GHz-frequency regime reported in D20 (see Fig.~\ref{fig:mrmoose} upper right). This spectral break (as well as the low-frequency turnover) may constrain the jet properties, although without a definitive redshift, we postpone this analysis to a future publication.}
We report the monochromatic luminosity at different rest-frame frequencies (500\,MHz, 3\,GHz and 100\,GHz) in Fig.~\ref{fig:money_plot}b. When compared to the HzRG selection limit \citep[L$_{\rm 3\,GHz}>10^{26}$\,W\,Hz$^{-1}$;][]{seymour_massive_2007}, \oddball\, meets the HzRG criterion in terms of radio luminosity \added{at $z\gtrsim1$}. The dashed part of the 500-MHz luminosity track indicates where we extrapolate from our best SED fit as the observed\added{-frame} frequency shifts outside of the MWA frequency coverage ($\nu_{\rm obs}<70\,$MHz). 

As the redshift of the source increases, the rest-frame frequency of the ALMA Band 3 image enters the sub-mm regime where dust may contribute significantly (the highest-frequency data may suggest an upturn; see Table~\ref{tab:radio_flux}). Given the negative $k$-correction of the cold gas in galaxies \citep{blain_history_1999} and the increase in CMB temperature with redshift, the dust contribution significantly affects the SED \citep{da_cunha_effect_2013}. We can use this to constrain the redshift range. 

While a significant contribution from dust is ruled out from our SED fitting (see Section~\ref{sec:radio_SED}), we present, as a blue shaded area in Fig.~\ref{fig:money_plot}b, the contribution of a modified blackbody \added{at the restframe frequency of 100\,GHz} (see Eq.~\ref{eq:mbb}) with $M_{\rm dust}=$10$^{7}$\,M$_\odot$ (the limit provided by \mrmoose; see Fig.~\ref{fig:mrmoose}). The temperature range is fixed as $T_{\rm CMB}(z)<T<60\,$K, where the minimum allowed temperature at a given redshift is the corresponding CMB temperature. Therefore, the allowed range of dust temperatures decreases with redshift, which could allow one to effectively ``break'' the classical temperature--redshift degeneracy for the highest-redshift sources. However, we remind the reader that numerous approximations are used in some of the terms (such as the dust composition and grain size distribution).

%The key information to remember from this second panel:
Our key conclusions from the second panel are as follows.
\begin{itemize}
    \item For $z>1$ sources, $L_{\rm 3\,GHz}$ is the preferable measure, as the rest-frame luminosity calculation does not require an extrapolation.
    \item For $z>1$, \oddball\, is considered \deleted{as} a powerful radio galaxy, similar to the \citet{seymour_massive_2007} sample.
    \item If the source is at $z<2$, it has very little dust, $M_{\rm dust}<10^7$\msun, and/or a higher dust temperature, $T_{\rm dust}>20\,$K (Section~\ref{sec:radio_SED}).
    \item For $z<1$, the source has a radio luminosity \added{on par with} radio-luminous dwarf galaxies \citep[e.g. ][]{mezcua_radio_2019}.
\end{itemize}

%ASPECS molecular gas mass calculation from RJ. 

\subsection{Size properties}
\label{sec:size}

The source appears unresolved in most of the radio data,\added{and marginally resolved in the FIRST \citep[deconvolved size 1.2\,arcsec$\times$1.0\,arcsec, ][]{becker_first_1995} and VLASS image \citep[deconvolved size 0.94\,arcsec$\times$0.58\,arcsec, ][]{gordon_catalog_2020}. Moreover, \oddball\ is unresolved} in the $K_s-$band image, thereby putting constraints on the physical size of both the radio structure and host galaxy, respectively (see Fig.~\ref{fig:money_plot}c). Our best size constraint in the radio comes from the ALMA 100-GHz data ($<0.7\,$arcsec), putting a $<6$\,kpc projected linear size limit if \oddball\, is at $z>1$. While this indicates that the source is small at high frequencies, one should note that this might not be the case at lower frequencies as the radio galaxies often have extended emission at these frequencies (i.e. lobes with lower-energy electrons).

Interestingly, the IPS observations (see Section \ref{sec:IPS}) provide 
\replaced{additional size constraints at low frequency.}{an upper limit on the estimated angular size and some hints about the radio morphology of the source.} Our source scintillates partially: at the $\sim$50\% level. As discussed previously \added{in Section~\ref{sec:data}}, this can be interpreted as \added{(i) \deleted{as} a single source 0.6\,arcsec across, (ii)} half of the 162-MHz flux being emitted from a region more compact than $\sim$ \replaced{0.3}{0.5}\,arcsec, which represents $<3\,$kpc at $z>1$ \citep[see Fig. 6 in][]{chhetri_interplanetary_2018} \added{or, (iii) a double source separated by at least 0.3\,arcsec of which one is partially resolved at $\sim$0.3\,arcsec. Given that the radio size of our source is \replaced{$<1.2$\,arcsec}{unresolved} at at 1.4, 3 and 100\ghz~ (i.e. the VLASS, FIRST and ALMA \replaced{size}{resolution}), we can likely discard the third option as we should be able to see some partially resolved features \replaced{at}{in} these frequencies. Moreover, the two first hypotheses basically assume that at least half of the flux is in a structure larger than $\sim0.3$\,arcsec. This translates into a 3-10\,kpc region (assuming $z>1$ and the FIRST/VLASS \replaced{size}{resolution}) or 3-6\,kpc (using the ALMA resolution as a limit with the assumption that the 100\ghz\ continuum traces the same physical process), and the remaining fraction (if any) at $<3$\,kpc.} \deleted{Moreover, the remainder of the flux may form a more extended and complex structure, which we can conservatively estimate to be smaller than $\sim$1\,arcsec, (i.e. the VLASS, FIRST and ALMA resolution) as our source is still unresolved at 1.4\ghz, 3\ghz~and 100\ghz.} 

We can compare the \replaced{observed}{radio} size to the maximum size estimated to be achievable for radio sources using the modelling presented in \citet{saxena_modelling_2017}, which uses the \citet{kaiser_luminosity_2007} framework. For a high-redshift source, the inverse-Compton\footnote{\added{The} up-scattering of lower-energy CMB photons \replaced{to high energies (X-ray) by}{with the} relativistic electrons.} (iC) effect will be significant, reducing and constraining the radio emission, and explaining the smaller radio size. Interestingly, the size limit from our unresolved observations at different frequencies is compatible with an iC-limited size for \oddball.

The \ks-band image provides an upper limit of 0.8\,arcsec for the projected angular size of the host galaxy. When compared with the size evolution of star-forming galaxies \citep[dark red line in Fig.~\ref{fig:money_plot}c;][]{allen_size_2017}, this upper limit seems consistent for any solution $z>1$, albeit with caveats given the uncertainties on the galaxy type of \oddball\, and the extrapolation outside of the fitted redshift range (dashed part of the line). Conversely, the observed size would suggest a very small system at $z<1$, on scales of a few kpc or even smaller. \deleted{For the lowest part of the redshift range, it even questions the extragalactic nature of \oddball. }

We now summarise the key points from this section.
\begin{itemize}
\item The unresolved radio data indicates a compact source at \replaced{all}{most} frequencies\deleted{with a luminosity on par with the brightest radio sources in the sky at $z>1$.}
\item The IPS observations provide us with an upper limit for the angular scale within which half of the radio flux at 162\mhz\, is contained: \replaced{0.3}{0.5}\,arcsec, corresponding to $<3$\,kpc at $z>1$. \added{Moreover, there is possibly some extended structure out to $\sim$10\,kpc (the VLASS/FIRST resolution data), or $\sim$6\,kpc (assuming the ALMA data traces the same emission).}
\item For $z<1$, the \added{host galaxy} \deleted{system} would be unusually small given our \ks-band observation, which \added{would be unprecedented in nature for a galaxy (<1\,kpc in size).} \deleted{would question the nature of \oddball\, being an extragalactic source.}
\end{itemize}

%\warning{Maybe we can say something about bright sources $<10\,$kpc are  typically either blazars or GPS sources? I note that if there is a turnover at 100\,MHz and the sources is $<10\,$kpc then that is consistent with a high-z GPS source (O'Dea works passim). Maybe even plot the GPS upper size assuming the 100\,MHz turn-over is real?} - commented for now

\subsection{Molecular gas properties}
\label{sec:mol}

\begin{figure*}
    \centering
    \includegraphics[width=0.9\textwidth, trim= 0 0 200 0,clip]{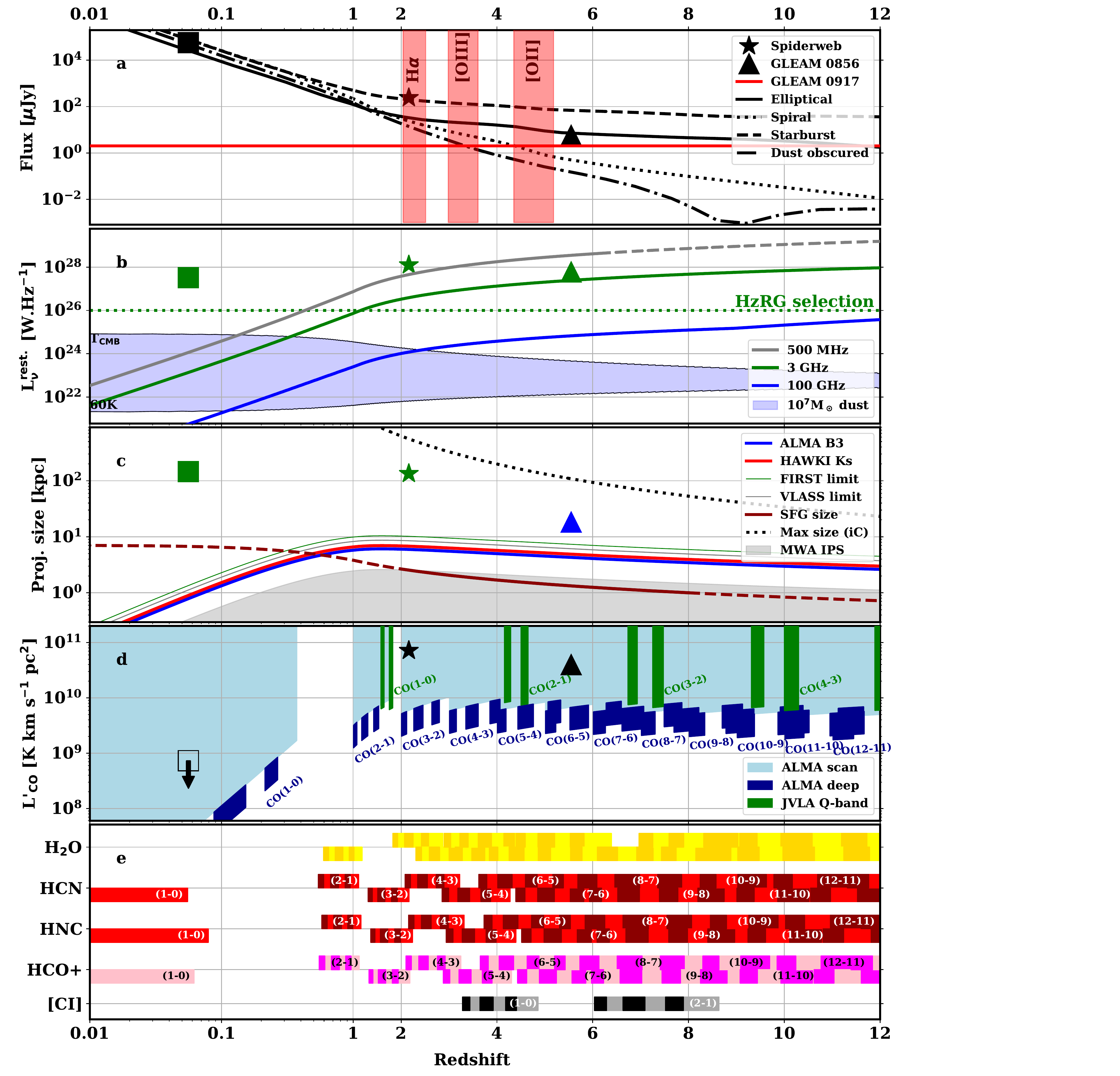}
    \caption{Figure synthesising the known constraints on \oddball\ as a function of redshift. Note that the redshift axis changes from a log to a linear scale at $z=1$. We present Cygnus A as a square (the open symbol in panel {\it (d)} indicating an upper limit), the Spiderweb galaxy as a star and GLEAM\,J0856+0224 as a triangle (see Section~\ref{sec:z_0917}). {\it From top to bottom:} $(a)$ \ks-band \added{flux density} as a red line) with \added{10$^{12}$\msun-normalised }stellar templates (see Section~\ref{sec:money_plot_near-IR} for more details), where the shaded light red areas represent potential contamination by  atomic lines; $(b)$ radio luminosities at different rest-frame frequencies (solid lines), with the dashed line showing the extrapolation beyond what we can constrain in the observed-frame, the blue shaded area representing a 10$^7$\,\msun~blackbody with a range of allowed temperatures (see Section~\ref{sec:radio_SED}), and the green dotted line showing the HzRG selection limit from \citet{seymour_massive_2007}; $(c)$ projected size at different frequencies along with (i) the star-forming galaxy size evolution (solid dark red line) from \citet[][]{allen_size_2017} (note the extrapolation as a dashed line), (ii) the relevant survey resolutions, (iii) the IPS size domain as a grey shaded area (see Section~\ref{sec:size}), and (iv) the maximal iC size from \citet{saxena_search_2018}; $(d)$ the accessible part of the \lpco--$z$ parameter space (shaded regions) given the  sensitivities of our observations (see Section~\ref{sec:mol} for more details); and $(e)$ supplementary molecular lines, with their observable ranges and the respective sensitivities from the ALMA spectra (see Section~\ref{sec:mol}).}
    \label{fig:money_plot}
\end{figure*}

Despite the non-detection of emission lines in our VLA and ALMA spectra, we can estimate useful upper limits \added{on the molecular gas properties} given the sizeable integration time spent on the source (Figs.~\ref{fig:money_plot}d and~\ref{fig:money_plot}e). We use the following formula to derive a 3$\sigma$ upper limit on the strength of each \added{CO} line \citep[adapted from Eq. 153 of][]{meyer_tracing_2017}: 

\begin{equation}
I_{\rm CO}<3\sigma\Delta\nu \sqrt{\frac{{\rm FWHM}}{\Delta\nu}},
\end{equation}

{\noindent where} $I_{\rm CO}$ is the integrated line flux in \icounit, $\sigma$ is the noise of the channel in Jy, $\Delta_\nu$ is the channel width in \kms, and FWHM is the full width at half maximum in \kms. Assuming FWHM=750\kms\ \added{(roughly three 80-\mhz-channels}\footnote{Some variation of the detection limit is expected when assuming a different line width, as it will correspond to a different channel width.}\added{) which is typical for high-redshift quasars and HzRGs} \citep[e.g.][]{carilli_cool_2013}, this leads to $I_{\rm CO}^{\rm ALMA}<0.09\,$\icounit and $I_{\rm CO}^{\rm VLA}<0.04\,$\icounit. \added{Note that this formula can be applied to other lines, such as [CI], and will lead to similar upper limits if assuming the same line width. Note that we use the radio convention for frequency--velocity conversion.}

From the integrated line flux limit, it is now possible to calculate a CO line luminosity ($L'_{\rm CO}$) limit, with the classical formula from \cite{solomon_molecular_2005}:
\begin{equation}
\small
L'_{\rm CO} = 3.25 \times 10^7 \left( \frac{I_{\rm CO}}{{\rm Jy\,km\,s^{-1} }} \right) \left( \frac{D_{L}}{\rm Mpc} \right)^2 \left( \frac{\nu_{\rm rest}}{{\rm GHz}} \right)^{-2} (1+z)^{-1},
\end{equation}
where $D_L$ is the luminosity distance and $\nu_{\rm rest}$ is the line rest-frame frequency. 

Fig.~\ref{fig:money_plot}d presents the accessible parameter space from our respective limits with ALMA (light and dark blue shaded areas) and VLA (green shaded area), for the corresponding line and redshift. The discontinuous shape for the deeper ALMA observations is due to the narrower frequency coverage compared to the original spectrum. \deleted{For the sake of simplicity, we assume the same line width of 750\kms\, for the line (corresponding to three 80-MHz channels in the ALMA data), in the expected range for powerful radio galaxies \mbox{\citep{carilli_cool_2013}}}. We note that we cannot access some parts of the parameter space: the $z<1$ and $1.8<z<2$ regions \citep[an observational constraint; see][]{weis_alma_2013}, as well as fainter systems with  \lpco$<10^{9.5}$\lpcounit. However, we do have good coverage for the rest of the redshift solutions for intrinsically bright CO emission \citep[\lpco $>10^{9.5}$\lpcounit, where most of the powerful systems are detected;][]{carilli_cool_2013}. 

A limitation not shown in  Fig.~\ref{fig:money_plot}d (that is very hard to assess without the redshift of the source) is the potential disappearance of the CO lines at the highest-redshift end. This would be due to the decreasing contrast of the line emission with the background light, either from the strong and compact radio emission from the source itself (creating a mix of emission and absorption of the CO lines) or the CMB. Indeed, as the CMB temperature increases with redshift, the gas floor temperature will be locked in with the background radiation and therefore the lower CO transitions will be unobservable \citep[e.g.][]{zhang_gone_2016}. This phenomenon particularly affects the lower rotational CO transition lines given the gas temperature, i.e. the VLA observations.

Finally, in Fig.~\ref{fig:money_plot}e, we report other bright molecular lines \added{transitions} in the ALMA frequency range, \added{potentially reaching above mJy level, and therefore detectable by ALMA}. We do not report their predicted intensities as (i) no observations are readily available for HzRGs \added{for the whole redshift range, (ii) while the lower HCN/HCO+ transitions would be detected \citep[$z<1$; e.g.][]{canameras_plancks_2021}, the higher ones are very likely to be well below our detection limit given that \oddball\ does not appear to be a lensed source} \cite[][]{riechers_dense_2010,spilker_rest-frame_2014}, and (iii) the special case of the H$_2$O molecule requires a complex set of assumptions/calculations \citep[e.g.][]{van_der_werf_water_2011,yang_water_2013} but some bright transitions are detected in high-redshift systems \citep[e.g.][]{weis_alma_2013,wang_alma_2013,gullberg_alma_2016,lehnert_etching_2020}. We draw attention to the [CI] lines, which cover the ranges $3<z<5$ and $6<z<9$. We expect these lines to have similar fluxes to the adjacent CO lines \citep[e.g.][]{gullberg_alma_2016}, which therefore put further constraints on the likely redshift of \oddball. 

The key implications of Figs.~\ref{fig:money_plot}d and \ref{fig:money_plot}e are as follows.
\begin{itemize}
    \item The redshift ranges $0.4<z<1.0$ and $1.8<z<2.0$ are not covered with our technique when solely considering the CO lines, and we could be missing a bright molecular gas source. 
    \item When taking into account secondary lines (HCO+, HCN/HNC and H$_2$O), the redshift window that is not covered possibly narrows to $0.4<z<0.6$,  favouring a higher-redshift solution and much fainter systems.
    \item The source is likely a molecular gas-poor system, especially for the higher-redshift solutions \deleted{where the galaxy is expected to be massive and young, i.e. forming stars.}
    \item The VLA data do not add strong redshift constraints as they are comparatively shallower than the ALMA data. These VLA data may also be affected by the aforementioned CMB effect at the high-redshift end.
\end{itemize}

%\section{Discussion}
\section{Constraining the redshift of \oddball}
\label{sec:disc}

In this section, we discuss the nature and the likely redshift of \oddball, distilling the information presented in \S\ref{sec:z_0917}. Firstly, we consider the hypothesis that \oddball\, is a Galactic source. The most obvious candidate in this case would be a pulsar given: (i) its faintness in \ks-band \citep[where few pulsars have been detected; e.g.][]{mignani_near-infrared_2012}, (ii) its extreme radio/near-IR flux density ratio (see Fig.~\ref{fig:Krad}), (iii) the compact nature of the source, (Fig.~\ref{fig:money_plot}) and, (iv) the low-frequency spectral index, similar to detected pulsars \citep[e.g.][]{murphy_low-frequency_2017}.  However, we argue against this possibility for \oddball\, for the following reasons: (i) none of the radio data considered in Section~\ref{sec:polarisation} have a strong polarised component up to 10\ghz, (ii) it does not appear to be coincident with a pulsar in currently available catalogues\footnote{The nearest pulsar to \oddball\ in the ATNF Pulsar Catalogue is at an angular distance of 7\degree; see www.atnf.csiro.au/research/pulsar/psrcat/ \citep{manchester_australia_2005}. Furthermore, \oddball\, is not detected in MWA pulsar surveys to date (R. Bhat, private communication).}, (iii) there is no evidence in the radio spectrum of variability, which we might expect from a pulsar, particularly when the spectrum has been compiled using data from a variety of observations taken at different epochs, and (iv) the NSI value (Section~\ref{sec:IPS}) would require a contrived scenario such as a small pulsar wind nebula or a coincidental alignment of a compact radio source with a steep-spectrum pulsar, and is also incompatible with a lone pulsar (which have an \replaced{NSI$>$0.9}{high NSI}).

%Although \oddball is not associated with a known pulsar\footnote{There are no known pulsars with in 1\,degree in the ATNF Pulsar Catalogue: www.atnf.csiro.au/research/pulsar/psrcat/. Also it is not detected in MWA pulsars surveys to date (R. Bhat, private communication).}, compact radio sources with steep spectra and significant polarisation have been followed up and subsequently found to be Galactic pulsars in previous studies \citep[e.g.][]{navarro_very_1995}. Furthermore, there is no evidence in the radio spectrum of variability, which we might expect from a pulsar, particularly when the spectrum has been compiled using data from a variety of observations taken at different epochs. Additionally, the optical and near-IR (near-IR) properties do not particularly suggest a pulsar origin, although, for example, some pulsars have been detected at near-IR wavelengths before \citep[e.g.][]{mignani_near-infrared_2012}. Hence this combination of radio and near-IR properties does not provide compelling evidence at all that \oddball is a nearby pulsar.

As a galaxy at $z<0.4$, \oddball\, would have very peculiar properties \added{for a radio-loud AGN or a compact galaxy, as suggested by the radio emission. These are:} (i) a low stellar mass, \mstel$<10^9$\msun\,(Fig.~\ref{fig:money_plot}a; note that this is in tension with the stellar mass of Cygnus A); (ii) very molecular gas-poor and with a low dust content (see Fig.~\ref{fig:money_plot}b,d); (iii) an extreme radio-to-near-IR flux density ratio, which is similar only to Cygnus A in the local Universe (see Fig.~\ref{fig:Krad}); \deleted{and} (iv) a very small size, $<1$\,kpc (Fig.~\ref{fig:money_plot}c); \added{and (v) a very obscured optical-near-IR SED (Fig.~\ref{fig:optical_SED}), in tension with (ii)}. For all these reasons, we consider this redshift range as unlikely.

For the $0.4<z<1.0$ range, \oddball\, has somewhat the same range of properties as described above, but less extreme: the mass is higher and the size is larger. We cannot be certain that \oddball\, does lie in this redshift range as the ALMA data do not cover any CO lines. However, Fig.~\ref{fig:optical_SED} seems to be inconsistent with this solution \added{except for a very obscured object ($A_{\rm V}>4$)}. \deleted{For a faint \ks-band detection and a non-detection in the optical, a very red object is required, which would be indicative of dust.} Yet, the SED at longer wavelengths (Fig.~\ref{fig:mrmoose}) does not suggest a large amount of dust ($M_{\rm dust}<10^{7}$\msun). Moreover, there is an overlap with the HCO$+$(2-1), HCN(2-1) and H$_2$O lines in this redshift range, decreasing the possibility of this redshift range even further \added{(which could be detected if reaching the mJy level)}. We also consider this solution unlikely.

At $z>1$, the effects due to redshift on parameters such as size and the CO luminosity detection limit become nearly constant as a result of cosmological effects. Also, the \ks-band flux indicates a massive system (\mstel$>10^{11}$\msun) of $<$8\,kpc in diameter. Given the radio luminosity, this source is definitely compatible with being a powerful radio galaxy (Fig.~\ref{fig:money_plot}). Even the radio to near-IR flux density ratio, albeit on the higher end of the distribution, is compatible (Fig.~\ref{fig:Krad}). The main differences compared with previous $1<z<5$ samples of powerful radio galaxies \citep[][]{de_breuck_spitzer_2010} are that \oddball~ is a relatively gas-poor system and is at the smaller end of the radio size distribution \added{($<0.8$\,arcsec; unresolved at 100GHz)}. Note \added{that there is a second gap for CO lines in the ALMA coverage at $1.8<z<2$.} \deleted{the $1.8<z<2$ range, where we have our second ALMA data gap}. However, the optical to near-IR SED does not suggest an object in this redshift range, for the same reasons elucidated in the previous paragraph. 

\added{The $z>2$ range marks the beginning of the redshift range where the ALMA data provides us with at minimum one CO line. Note that} for the \replaced{$2<z<3$}{$2<z<6$} range, only a dust-obscured source (with $A_{\rm V}\sim3.5$) is possible given the optical and near-IR photometry (see Fig.~\ref{fig:optical_SED}). \added{This would push \oddball\ into the infrared luminous galaxies regime.} A significant amount of dust has two immediate corollaries: (i) by absorbing the UV--optical light, the dust would re-emit in the far-IR \added{(into the {\it Herschel} coverage)}, but the ALMA continuum provides us with a tight constraint (see Fig.~\ref{fig:mrmoose}), and (ii) a order of magnitude estimate of the gas mass from the dust mass assuming a conservative gas-to-dust mass ratio ($>100$) indicates that the CO(2-1) line enters our detectable range \citep[with $\alpha_{\rm CO}=0.8$ and CO line ratios from][]{carilli_cool_2013}. \added{Finally, the very specific case of a  thin dust lane localised on the line-of-sight, which could reproduce the strong obscuration seen in optical/near-IR and the lack of large far-IR/submm contribution cannot be fully excluded with our present data, but appears much less likely.}

%\added{From the insert in the lower left of Fig.~\ref{fig:optical_SED}. There is a degeneracy between the acceptable redshift and dust attenuation ($A_v$). Given extreme dust extinction ($A_V>4$) is unlikely, the most probable solution in this redshift range is $z\sim 3$ and $A_v\sim 3.5$.}
\deleted{There are two possible scenarios: a heavily obscured ($A_V>5$) object at low redshift ($z<2$), or a moderately obscured ($A_V\sim4$) object at intermediate redshift $2<z<6$. We consider the first scenario unlikely, as a significant amount of dust has two immediate corollaries: (i) by absorbing the UV--optical light, the dust would re-emit in the far-IR, but the ALMA continuum provides us with a tight constraint (see Fig.~\ref{fig:mrmoose}), and (ii) a order of magnitude estimate of the gas mass from the dust mass assuming a conservative gas-to-dust mass ratio ($>100$) indicates that the CO(2-1) line enters our detectable range [....]. As for the second hypothesis, the range of possible extinction is very restricted, with $A_V\sim4$. Moreover, for the higher-redshift range ($z>3$), the \ks-band flux shows an increasing tension with the observed stellar mass range for HzRGs (Fig.~\ref{fig:money_plot}a). This basically reduces the dust-obscured source to having a very specific set of properties, with $A_V\sim4$ and a $2<z<3$ range. While not fully impossible, we argue that these solutions are unlikely.}

We now explore the \replaced{$z>7$}{$z>5$} range, which could explain the optical to near-IR SED due to the sharp drop of the Lyman continuum resulting from IGM absorption. One should note that some dust extinction could lower this redshift limit by producing a redder SED and still be consistent with the upper limits. However it cannot be too red or it would be incompatible with the WISE data (see Section~\ref{sec:money_plot_near-IR}). A very high redshift also seems to be consistent with the extreme radio-to-near-IR flux density ratio, given the trend from Fig.~\ref{fig:Krad}. Moreover, the extreme ratio and the faintness in $K_s$-band implies a very luminous radio source, similar to powerful radio galaxies in the $1<z<5$ range. The mass of this system would be on the higher end of the distribution for HzRGs, approaching the $10^{12}$\msun\ limit from  \citet[e.g.][]{rocca-volmerange_radio_2004}. A spiral scenario would only be able to reproduce the \ks-band flux density with an unrealistic mass (\mstel$>10^{14}$\msun; see dotted line in Fig.~\ref{fig:money_plot}a). A pure starburst could reproduce the \ks-band flux density, at an equivalent lower mass due to the higher light-to-mass ratio of the younger and more massive stars. Yet once again, a massive starburst (of the order of $10^{10}$\msun) would involve a significant amount of dust and gas as well as a possible disturbed morphology (notably in rest-frame UV / observed-frame near-IR). While the near-IR size could remain consistent with a star-forming galaxy (see dark line in Fig.~\ref{fig:money_plot}c and the complete optical/near-IR SED in Fig.~\ref{fig:optical_SED}, it suggests a more passively evolving galaxy, and thus a flatter template in the UV. As for the molecular gas content, \oddball\, would have a value $<10^{9.5}$\lpcounit, which, while probably lower than one would expect, is consistent with the little amount of dust (if any) from the radio to far-IR SED (Fig.~\ref{fig:mrmoose}). Also, \added{it is important to} keep in mind that at very high redshifts, \ks-band corresponds roughly to the UV regime, so any presence of dust would have a strong effect on the continuum. 

\deleted{As such, we favour the very high-redshift solution for \oddball, assuming a conservative lower limit of $z>5$ from the optical SED (Fig.~\ref{fig:optical_SED}) if some low level of dust extinction is accounted for (\replaced{up to $A_{\rm V}=2$}{$A_{\rm V}<1$}\deleted{ with a non-starburst template)}, or $z \gtrsim7$ if dust extinction is non-existent \added{or limited}. Interestingly, all available evidence from the data we have seems to place this source at $z>5$, which would make the selection technique from D20 very efficient (50\%, albeit with small number statistics) for finding powerful radio galaxies at the end of the Epoch of Reionisation.}

\section{CONCLUSION}
\label{sec:conc}

We presented new ALMA and VLA follow-up data obtained in order to explore the putative $z=10.15$ nature of \oddball\, from D20. The deeper observations do not confirm the detection of the low signal-to-noise lines observed in the first ALMA spectrum, therefore ruling out this redshift solution. \added{Adding multi-wavelength public imaging data from optical to infrared (HSC, {\it WISE} and {\it Herschel}) and additional information from radio frequencies (IPS and polarisation), we are able to narrow the properties of \oddball\ significantly. In particular, the compactness both in near-IR and radio, the implied low amount of dust and molecular gas, the large radio luminosity and the extreme radio-to-near-IR flux density ratio, leads to a very peculiar source for any low-$z$ solutions.} We argue that a \replaced{$z>7$}{$z>5$} solution is more likely, \added{with a possible, albeit much less likely, solution at $2<z<3$ in case of a peculiar dust geometry and extreme obscuration ($A_{\rm V}\sim3.5$) to reproduce the optical/near-IR data, and the non detection of cold dust continuum}. Our options are now near-IR spectroscopy and additional ALMA scans at a different frequency range to find the [CI] or [CII] lines.
%The hunt continues. 

\begin{acknowledgements}
The authors would like to thank Cathryn Trott for useful discussion on the IGM properties, Simon Driver and Luke Davies for their help to accessing the GAMA dataset, and Chris Riseley for providing the upper limit from the POGS survey. JA acknowledges financial support from the Science and Technology Foundation (FCT, Portugal) through research grants PTDC/FIS-AST/29245/2017, UIDB/04434/2020 and UIDP/04434/2020. \added{We also thank the anonymous referee for the useful suggestions, helping in clarifying this manuscript.}

The National Radio Astronomy Observatory is a facility of the National Science Foundation operated under cooperative agreement by Associated Universities, Inc.

This paper makes use of the following ALMA data: ADS/JAO.ALMA\#2019.A.00023.S,2017.1.00719.S. ALMA is a partnership of ESO (representing its member states), NSF (USA) and NINS (Japan), together with NRC (Canada), MOST and ASIAA (Taiwan), and KASI (Republic of Korea), in cooperation with the Republic of Chile. The Joint ALMA Observatory is operated by ESO, AUI/NRAO and NAOJ.

This publication makes use of data products from the {\it Wide-field Infrared Survey Explorer}, which is a joint project of the University of California, Los Angeles, and the Jet Propulsion Laboratory/California Institute of Technology, funded by the National Aeronautics and Space Administration.

Based on data collected at the Subaru Telescope and retrieved from the HSC data archive system, which is operated by Subaru Telescope and Astronomy Data Center at National Astronomical Observatory of Japan.

The Hyper Suprime-Cam (HSC) collaboration includes the astronomical communities of Japan and Taiwan, and Princeton University. The HSC instrumentation and software were developed by the National Astronomical Observatory of Japan (NAOJ), the Kavli Institute for the Physics and Mathematics of the Universe (Kavli IPMU), the University of Tokyo, the High Energy Accelerator Research Organization (KEK), the Academia Sinica Institute for Astronomy and Astrophysics in Taiwan (ASIAA), and Princeton University. Funding was contributed by the FIRST program from Japanese Cabinet Office, the Ministry of Education, Culture, Sports, Science and Technology (MEXT), the Japan Society for the Promotion of Science (JSPS), Japan Science and Technology Agency (JST), the Toray Science Foundation, NAOJ, Kavli IPMU, KEK, ASIAA, and Princeton University. 

This paper makes use of software developed for the Large Synoptic Survey Telescope. We thank the LSST Project for making their code available as free software at http://dm.lsst.org

The Pan-STARRS1 Surveys (PS1) have been made possible through contributions of the Institute for Astronomy, the University of Hawaii, the Pan-STARRS Project Office, the Max-Planck Society and its participating institutes, the Max Planck Institute for Astronomy, Heidelberg and the Max Planck Institute for Extraterrestrial Physics, Garching, The Johns Hopkins University, Durham University, the University of Edinburgh, Queen’s University Belfast, the Harvard-Smithsonian Center for Astrophysics, the Las Cumbres Observatory Global Telescope Network Incorporated, the National Central University of Taiwan, the Space Telescope Science Institute, the National Aeronautics and Space Administration under Grant No. NNX08AR22G issued through the Planetary Science Division of the NASA Science Mission Directorate, the National Science Foundation under Grant No. AST-1238877, the University of Maryland, and Eotvos Lorand University (ELTE) and the Los Alamos National Laboratory.

Based in part on data collected at the Subaru Telescope and retrieved from the HSC data archive system, which is operated by Subaru Telescope and Astronomy Data Center at National Astronomical Observatory of Japan.

The Australia Telescope Compact Array is part of the Australia Telescope National Facility which is funded by the Australian Government for operation as a National Facility managed by CSIRO. We acknowledge the Gomeroi people as the traditional owners of the Observatory site.

This research has made use of the NASA/IPAC Extragalactic Database (NED), which is operated by the Jet Propulsion Laboratory, California Institute of Technology, under contract with the National Aeronautics and Space Administration.
\end{acknowledgements}

\bibliographystyle{pasa-mnras}
%\bibliography{1r_lamboo_notes}
\bibliography{references.bib}

\end{document}